\documentclass[journal]{IEEEtran}

\usepackage{amsmath}
\usepackage{amssymb}
\usepackage{amsfonts}
\usepackage{graphicx,graphics,color,psfrag}
\usepackage{epsfig}
\usepackage{psfrag}
\usepackage{cite}
\usepackage{latexsym}
\usepackage{url}
\usepackage{subcaption}
\usepackage{booktabs}
\usepackage{color}
\usepackage{multirow}
\usepackage{mathtools}
\usepackage{bm}
\usepackage{booktabs}
\usepackage{algorithm}
\usepackage{longtable}
\usepackage{tikz}
\usetikzlibrary{arrows.meta,positioning,shapes.geometric,fit,calc,backgrounds}
\usepackage{enumitem}
\setlist[itemize]{leftmargin=*, topsep=1pt, itemsep=1pt, parsep=0pt, partopsep=0pt}
\usepackage{algpseudocode}
\usepackage{bm}
\usepackage{array}
\usepackage{verbatim}
\usepackage{indentfirst}
\usepackage{hyperref}
\usepackage{multirow}
\usepackage{bbm}
\usepackage{caption}
\graphicspath{{fig/}}

\PassOptionsToPackage{bookmarks={false}}{hyperref}
\IEEEoverridecommandlockouts

\allowdisplaybreaks[4] 
\newcommand{\revise}[1]{#1}
\newcommand{\bluechange}{\textcolor{black}}
\begin{document}
\title{Toward Intelligent Skies: Signal Processing and AI Foundations of Low-Altitude Wireless Networks}
\author{Weijie Yuan,~\IEEEmembership{Senior Member,~IEEE,} Geng Sun,~\IEEEmembership{Senior Member,~IEEE,} Jiacheng Wang,~\IEEEmembership{Member,~IEEE,} Jun Wu, Yuanhao Cui,~\IEEEmembership{Member,~IEEE,} Jiahui Li,~\IEEEmembership{Member,~IEEE,} Wei Zhang,~\IEEEmembership{Fellow,~IEEE,} George K. Karagiannidis,~\IEEEmembership{Fellow,~IEEE,}  Sumei Sun, \IEEEmembership{Fellow,~IEEE,} and Yonina C. Eldar, \IEEEmembership{Fellow,~IEEE}
\thanks{W. Yuan, J. Wu, and Y. Cui are with the School of Automation and Intelligent Manufacturing, Southern University of Science and Technology, Shenzhen 518055, China. (e-mail: yuanwj@sustech.edu.cn, wuj2021@mail.sustech.edu.cn, yuanhao.cui@bupt.edu.cn)}
\thanks{G. Sun and J. Li are with the College of Computer Science and Technology, Jilin University, Changchun 130012, China (e-mail: sungeng@jlu.edu.cn, lijiahui@jlu.edu.cn)}
\thanks{
J. Wang is with the College of Computing and Data Science, Nanyang Technological University, Singapore 639798 (email: jiacheng.wang@ntu.edu.sg)}
\thanks{W. Zhang is with the School of Electrical Engineering and Telecommunications, University of New South Wales, Sydney, NSW 2052, Australia (e-mail: w.zhang@unsw.edu.au)}
\thanks{G. K. Karagiannidis is with the Department of Electrical and Computer Engineering, Aristotle University of Thessaloniki, 54124 Thessaloniki, Greece (e-mail: geokarag@auth.gr)}
\thanks{S. Sun is with the Institute for Infocomm Research, Agency for Science, Technology and Research (A*STAR), Singapore 138632 (e-mail: sunsm@a-star.edu.sg)}
\thanks{Y. C. Eldar is with the Department of Electrical and Computer Engineering,
Northeastern University, MA 02115, USA (e-mail: y.eldar@northeastern.edu)}
}
\maketitle
\begin{abstract}
The rapid growth of low-altitude aerial services and applications, driven by uncrewed aerial vehicles (UAVs), calls for a new class of digital infrastructure beyond conventional terrestrial networks. The low-altitude wireless network (LAWN) has been proposed as dynamically reconfigurable three-dimensional architectures that integrate aerial and ground nodes to provide connectivity, sensing, and control in open, safety-critical airspace. This tutorial presents a comprehensive treatment of LAWNs from the joint perspectives of artificial intelligence (AI) and signal processing. We first review the historical evolution and architectural foundations of LAWNs, introducing altitude-based layers and functional planes, and summarizing the regulatory and standardization landscape. Building on this system view, we then discuss signal processing fundamentals for LAWNs, including 3D channel and system models, performance metrics, waveform and receiver design, localization and tracking, and multi-functionality co-design. Next, we survey AI techniques for LAWNs, covering discriminative and generative models for perception, control, resource management, and security, as well as emerging paradigms such as foundation models, large language models, and digital twins for mission planning and closed-loop optimization. To illustrate AI-signal processing integration in practice, we provide a case study of an AI-driven multi-tier LAWN with hybrid satellite, high-altitude, and ground nodes. The tutorial concludes by outlining key research challenges in architecture design, signal processing-AI co-design, safety and security, experimentation, and standardization, and by highlighting opportunities for LAWNs to evolve into dependable, AI-native infrastructure for the intelligent skies.
\end{abstract}
\newcolumntype{p}[1]{>{\centering\arraybackslash}m{#1}}

\section{Introduction}
\subsection{Motivations of the LAWN}
The proliferation of uncrewed aerial vehicles (UAVs) has transformed low-altitude airspace (typically below $3,000$ meters above the ground) into a dynamic arena for innovative applications, giving rise to what is increasingly termed the {low-altitude economy.} This emerging sector leverages UAVs for efficient, autonomous operations across diverse fields such as logistics, precision agriculture, public safety, environmental monitoring, and urban air mobility (UAM). For instance, UAVs now support rapid delivery services in e-commerce, real-time crop health assessment, disaster response coordination, and even passenger transport via electric vertical takeoff and landing (eVTOL) vehicles. The global drone market is projected to grow from approximately \$42 billion in 2025 to nearly \$90 billion by 2030\cite{MordorIntelligence2025}, reflecting a compound annual growth rate of 13-14\%. As for the broader {low-altitude economy}, it is expected to reach \$150-200 billion globally by 2030 and is poised for rapid expansion \cite{GrapheneRich2025}.

Conventional terrestrial networks built on static infrastructure struggle to meet these demands {in a 3D}, rapidly changing low-altitude airspace. This gap spans sensing, communication, and control, and is magnified by multi-objective tradeoffs in timeliness, energy, and scalability, as well as heightened security risks in open airspace. To this end, a low-altitude wireless network (LAWN) is expected to serve as the foundational infrastructure for this vision, enabling seamless connectivity, real-time data exchange, and coordinated control among aerial drones, ground stations, and other entities\cite{yuan2025ground}. Defined as a dynamically reconfigurable {(3D)} network that integrates aerial and terrestrial nodes, the LAWN supports high-speed data transmission, ultra-reliable low-latency command-and-control, and precise environmental sensing. Unlike traditional terrestrial or satellite networks, LAWNs address the distinctive demands of low-altitude environments, including high mobility, fluctuating network topologies, variable line-of-sight (LoS) propagation influenced by urban structures and weather, as well as stringent latency and reliability requirements.

The motivation for developing and advancing LAWNs is further reinforced by evolving regulatory frameworks worldwide. \revise{In the United States, the Federal Aviation Administration (FAA) has proposed rules to normalize beyond-visual-line-of-sight (BVLOS) operations, restricting them to altitudes at or below $400$ feet while emphasizing enhanced safety measures, aircraft equipage, and security policies to enable scalable commercial and public applications\cite{FAA2025}.} In China, the Civil Aviation Administration of China (CAAC) has formally integrated provisions for the low-altitude economy into its revised civil aviation law in 2025, becoming the first nation to do so \cite{BusinessAviation2025}. Similarly, the European Union Aviation Safety Agency (EASA) maintains regulations for structured low-altitude operations and traffic management to prioritize public safety \cite{EASA2025}. In Japan, drones weighing $100$ grams or more must be registered with a remote ID system, with two-tier pilot licensing for advanced operations, including BVLOS under specific approvals\cite{MLIT2025}. \revise{In Australia, the Civil Aviation Safety Authority (CASA) mandates registration for drones over $250$ grams and emphasizes remote ID compliance for BVLOS flights in controlled airspace\cite{CASA2025}.} Collectively, these evolving policies not only constrain unsafe operations but also create the conditions for dense, routine low-altitude services, thereby highlighting the need for dependable wireless infrastructures that can support compliant, scalable low-altitude ecosystems. In Table~\ref{table1}, we summarize existing efforts from governments and regulators relevant to LAWNs.

\begin{table*}[!ht]
\centering
\setlength{\tabcolsep}{3pt}
\renewcommand{\arraystretch}{1.18}
\caption{Government/Regulatory Efforts Relevant to the LAWN}
\label{table1}
\begin{tabular}{|p{2.2cm}|p{3.8cm}|p{4.0cm}|p{6.0cm}|}
\hline
\textbf{Country/Region} & \textbf{Regulator} & \textbf{Programme/Instrument} & \textbf{Main Focus Areas} \\
\hline
Global & International Civil Aviation Organization (ICAO)& Unmanned Aircraft System Traffic Management Framework (UTM Framework) \cite{ICAO_UTM2023}& Core UTM capabilities, roles/services, harmonization with Air Traffic Management (ATM); guidance for States deploying digital low-altitude services. \\
\hline
Global & International Telecommunication Union Radiocommunication Sector (ITU-R) & World Radiocommunication Conference 2023/2027 agenda \cite{itur2024future} & Spectrum for uncrewed aircraft systems (UAS) command-and-control; protection of aeronautical services; studies guiding bands relevant to LAWN communications. \\
\hline
Australia & Civil Aviation Safety Authority and Air-services Australia & Remotely Piloted Aircraft Systems and Advanced Air Mobility Strategic Regulatory Roadmap \cite{CASA2025}& 10--15 year integration plan for drones/AAM; risk-based rules and streamlined authorisations; FIMS as national UTM backbone and data-exchange hub. \\
\hline
Brazil & National Civil Aviation Agency and Department of Airspace Control & SARPAS Next Generation; proposed Brazilian Civil Aviation Regulation 100 \cite{Brazil_RBAC2025}& Registration/licensing and digital authorisations; transition to performance-/risk-based operations through updated regulation. \\
\hline
Canada & Transport Canada and Nav Canada& Remotely Piloted Aircraft Systems Traffic Management / RPAS Traffic Management \cite{Canada_RTM2020}& Concept of services for tracking, Remote ID, conflict detection/resolution; phased trials toward suburban/commercial BVLOS operations. \\
\hline
China & Civil Aviation Administration of China & Low-Altitude Economy (LAE) initiatives and general aviation steering measures \cite{BusinessAviation2025} & Regulatory system for LAE: airworthiness, flight operations oversight, flight-service platforms/dispatch, market regulation, infrastructure build-out and trials. \\
\hline
European Union & European Union Aviation Safety Agency and European Commission & U-space Regulations 2021/664, 2021/665, 2021/666 \cite{EASA2025} & U-space airspace/services: network Remote ID, strategic/tactical deconfliction, conformance monitoring; dynamic airspace reconfiguration and Member State implementation. \\
\hline
India & Directorate General of Civil Aviation & Drone Rules 2021; DigitalSky platform; National UTM Policy \cite{India_Drone2021}& Registration/UIN, licensing and permissions; DigitalSky governance and approval of UTM service providers; risk-based operational approach. \\
\hline
Japan & Ministry of Land, Infrastructure, Transport and Tourism and Japan Civil Aviation Bureau  & Level-4 Beyond Visual Line of Sight regime; national UAS Traffic Management steps \cite{MLIT2025} & BVLOS over populated areas via aircraft/operator approvals; phased UTM introduction, identification/registration, ecosystem standards coordination. \\
\hline
Korea & Ministry of Land, Infrastructure and Transport & Korea Urban Air Mobility Grand Challenge; K-Drone \cite{Korea_KUAM2025}& National demonstrations and regulatory sandbox; evaluation of safety risks (weather, birds, RF), urban corridors, and 5G-enabled low-altitude network concepts. \\
\hline
United Kingdom & UK Civil Aviation Authority & Civil Aviation Publication 722 series \cite{UK_CAP7222024}& Policy/guidance for UAS in UK airspace; operational categories, authorisations, design/airworthiness changes, airspace restrictions processes. \\\hline
United States & Federal Aviation Administration (FAA) & Beyond Visual Line of Sight rulemaking; BEYOND program; UAS Traffic Management Concept of Operations \cite{FAA2025} & Normalizing BVLOS operations, data-driven waivers and integration pilots; defining digital UTM services and roles. \\
\hline
United States & Federal Communications Commission& 5030--5091\,MHz Notice of Proposed Rulemaking on UAS command-and-control \cite{US_FCC2025}& Licensed spectrum framework for safety-critical UAS C2 links; coexistence/compatibility criteria (e.g., AeroMACS) and protection rules. \\
\hline
\end{tabular}
\end{table*}


Against the economic and regulatory background, the complexities of LAWNs, such as signaling and receiver design, multi-functionality co-design, as well as network security and privacy pose critical challenges for practical deployment. This necessitates advanced signal processing techniques for foundational tasks like target tracking, together with artificial intelligence (AI) for predictive analytics, autonomous decision-making, and adaptive learning from environmental data. By synergizing AI and signal processing, LAWNs can achieve substantial performance gains. This tutorial paper provides a comprehensive overview of signal processing and AI foundations of LAWNs, aiming to equip researchers and industry experts with the knowledge to push the intelligent skies forward.

\subsection{A Historical View of the LAWN: Multi-Disciplinary Convergence}

The evolution of the LAWN spans over 175 years, shaped by successive waves of military demand, technological innovation, and civilian adaptation\cite{newcome2004unmanned}. Early concepts emerged in the mid-19th century with balloons and basic remote actuation, followed by early-20th-century breakthroughs in autopilot and proportional control that made repeatable flight possible. The introduction of radio links added teleoperation, linking aerospace engineering with the nascent field of signal-based guidance. By the 1940s, radio-controlled targets such as the Radioplane demonstrated practical uncrewed flight and foreshadowed integration with robotics.

From the 1950s to 1970s, UAVs matured from prototypes to operational platforms. Radar detection, tracking, and radio-relay techniques advanced in parallel, marking the start of communication-assisted flight. In the 1970s–1990s, aerial photography, radar, and multispectral imaging from satellites and manned aircraft migrated to drones, giving rise to portable remote sensing for mapping and reconnaissance \cite{blom2010unmanned}. Simultaneously, solid-state electronics, GNSS, and lightweight inertial sensors enhanced navigation reliability, while computer vision and robotics introduced limited autonomy and environmental awareness.

The 2000s brought democratization. Lithium-polymer batteries, brushless motors, and compact IMUs enabled agile multi-rotors, while open autopilot platforms such as ArduPilot lowered the entry barrier for research and industry. UAVs found applications in inspection, agriculture, disaster response, and logistics. {Networking technologies, such as Wi-Fi, Bluetooth, UWB, and LoRa, enabled flexible telemetry}, command, and data relay across diverse distances and energy budgets.

The 2010s onward marked integration into communication infrastructures. LTE and 5G connectivity introduced aerial user equipment concepts, improving interference management and mobility. Onboard computing enabled real-time perception and planning, and civil regulations evolved through UTM and Remote ID frameworks\cite{wu2025low}.

%

\tikzset{
  >={Latex[length=1.6mm]},
  lawnfont/.style   = {font=\small},
  hub/.style        = {draw, rounded corners=4pt, very thick, align=center,
                       inner sep=5pt, minimum width=40mm, minimum height=12mm, fill=black!3},
  cluster/.style    = {draw, rounded corners=3pt, align=left, inner sep=4pt,
                       fill=black!1},
  slim/.style       = {draw, rounded corners=3pt, align=left, inner sep=3pt},
  edge/.style       = {very thick, -{Latex[length=2.5mm,width=1.5mm]}},
  edge2/.style      = {semithick, -{Latex[length=2mm,width=1mm]}},
}

\begin{figure*}[t]
\centering
\resizebox{\linewidth}{!}{%
\begin{tikzpicture}[lawnfont, node distance=7mm and 10mm]

\node[hub] (hub) {{\large LAWN}\\\scriptsize Coordinated, safety-managed 3D network};

\node[cluster, above left=13mm and 16mm of hub, anchor=south east, text width=45mm] (comm) {
\textbf{Communications Eng.}
\begin{itemize}
  \item “Cellular-in-the-sky”: A2G/A2A (LoS)
  \item 3GPP Rel-14/15: UAV as UE; C2 $\sim$100\,kbps, $<50$\,ms
  \item Rel-17 NR/mmWave; on-board RAN; massive MIMO
  \item URLLC; interference \& mobility mgmt.
\end{itemize}};

\node[cluster, above right=13mm and 16mm of hub, anchor=south west, text width=45mm] (csai) {
\textbf{Computer Science \& AI}
\begin{itemize}
  \item SLAM/VIO (GNSS-denied)
  \item CNN perception; semantic segmentation
  \item RL for trajectory/resource allocation
  \item Multi-agent RL; LLM supervisors; FL
\end{itemize}};

\node[cluster, below left=14mm and 16mm of hub, anchor=north east, text width=45mm] (robot) {
\textbf{Robotics}
\begin{itemize}
  \item Multirotors, VTOL hybrids; manipulators
  \item Fault-tolerant/\,$H_\infty$ control
  \item Modular payloads; cooperative lift
  \item Formation/coverage; collision avoidance
\end{itemize}};

\node[cluster, below right=14mm and 16mm of hub, anchor=north west, text width=45mm] (sp) {
\textbf{Signal Proc. \& Aero. Electronics}
\begin{itemize}
  \item Radar/EO/LiDAR fusion; KF/particle filters
  \item AOA/TDOA/UWB localization; spectrum sensing
  \item Adaptive beamforming; interference mitigation
  \item ISAC: shared waveforms (sense+comm)
\end{itemize}};

\node[slim, above=15.5mm of hub, text width=50mm] (goals)
  {\textbf{Service goals:} latency, availability, coverage probability, perception accuracy};

\node[slim, below=15.5mm of hub, text width=50mm] (energy)
  {\textbf{Energy \& safety in the loop:} battery-aware routing, health monitoring, charging/swap};

\node[slim, right=16mm of hub, text width=45mm] (twin)
  {\textbf{Airspace services:}
   \begin{itemize}
     \item Live assets/maps/energy/rules
     \item UTM/Remote ID/BVLOS; geofencing
     \item Mission scheduling \& airspace reservation
   \end{itemize}};

\begin{pgfonlayer}{background}
  \draw[edge] (comm) to[out=-20, in=160] (hub);
  \draw[edge] (csai) to[out=-160, in=20] (hub);
  \draw[edge] (robot) to[out=20, in=-160] (hub);
  \draw[edge] (sp)   to[out=160, in=-20] (hub);

  \draw[edge2] (goals.south)  to[out=-90, in=90]  (hub.north);
  \draw[edge2] (energy.north) to[out=90,  in=-90] (hub.south);
  \draw[edge2] (twin.west)    to[out=180, in=0]   (hub.east);
\end{pgfonlayer}
\end{tikzpicture}%
}
\caption{Multi-disciplinary convergence toward low-altitude wireless networks (LAWNs). LAWNs emerge from the coupling of (i) communications engineering, (ii) computer science and AI, (iii) robotics and control, and (iv) signal processing and aerospace electronics. This convergence is shaped by airspace services and energy–safety constraints, which jointly define system-level service goals such as latency, availability, coverage probability, and perception accuracy.}

\label{fig:lawn_convergence_ieee}
\end{figure*}
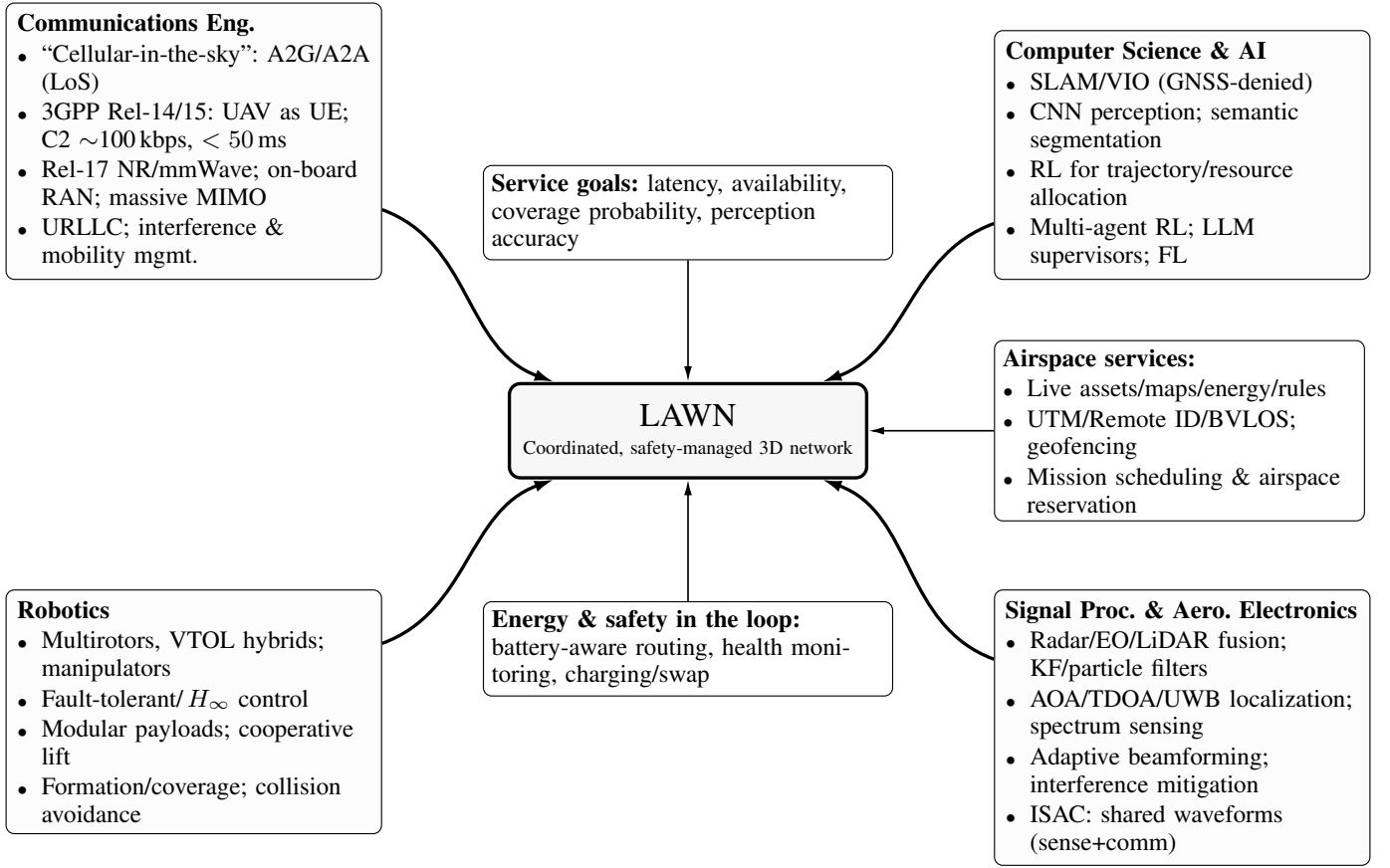

\subsubsection{Communications Engineering - The Cellular-in-the-Sky Paradigm}
The ``cellular-in-the-sky'' concept has been a core enabler of LAWNs by turning UAVs into networked nodes rather than isolated platforms. Early military systems in the 1960s mainly used drones as RF relays for C2\cite{barrow1970integrated}, but interference, short range, and crude resource control quickly exposed the limits of ad hoc aerial links. By the 1980s, satellite connectivity supported high-altitude platforms, and the rise of 2G/3G in the 1990s triggered interest in reusing terrestrial infrastructure to serve low-altitude UAV operations. In the 2000s, UAV–cellular integration gained momentum, especially for rapid coverage extension in remote or disaster-hit areas\cite{el2002cellular}. With 4G LTE, drones were increasingly treated as aerial user equipment (UE), and 3GPP Releases 14 and 15 formalized LTE support for UAVs, including beamforming-based interference mitigation, baseline C2 targets, and mobility handling for faster aerial users\cite{muruganathan2022overview}. More recently, 5G NR and beyond-5G efforts aim at URLLC-grade links and heterogeneous aerial networking for use cases such as UAM. Release 17 broadened options such as onboard radio access nodes and mmWave links for high-capacity A2A/A2G connectivity, while also highlighting issues like massive-MIMO pilot contamination. 

\subsubsection{Computer Science and AI – Agent UAVs and Computer Vision}
Computer science and AI have transformed UAVs from remotely piloted platforms into autonomous agents with onboard perception, decision-making, and cooperation. Early efforts in the 1950s–1960s relied on rule-based navigation. By the 1980s, computer vision enabled basic object/region recognition from edge features and pattern matching, establishing vision-based sensing for UAV missions\cite{besl1985three}. As fleets became practical, multi-agent systems modeled UAVs as distributed agents that coordinate under limited communication, leading to algorithms for formation, coverage, and task allocation. Since the 2000s, learning-based methods have strengthened autonomy: classical models (e.g., SVMs) supported obstacle and terrain classification, while SLAM fused visual and inertial data for localization and mapping in unknown environments\cite{caballero2009vision}. Deep learning, especially CNNs, enabled real-time detection, segmentation, and scene understanding on small UAVs, supporting navigation in cluttered spaces and infrastructure inspection. Reinforcement learning (RL) further optimized trajectories and communication policies under energy–latency–coverage tradeoffs, and multi-agent RL (MARL) extends this to swarms for sensing, relaying, and search-and-rescue\cite{liu2019reinforcement}. In the 2020s, agentic AI has become central to LAWN concepts\cite{sapkota2025uavs}. Generative models and LLMs can translate mission directives into task graphs and constraints, while federated learning supports distributed training without centralizing sensitive data. Vision has expanded to dense 3D mapping, reconstruction, and event-based sensing for low-latency maneuvers, supporting real-time situational awareness and more robust BVLOS operations and Remote ID compliance.

\subsubsection{Robotics and Control – Aerial Robots}
Robotics provides the physical and control foundations for LAWNs, enabling UAVs to function as agile aerial robots. Early milestones include the 1917 Kettering Bug and, by the 1960s, target drones using PID control for basic stabilization and navigation. Later advances added manipulators and lightweight sensing for inspection and interaction, with simple proximity sensors supporting obstacle avoidance. A major inflection came with multirotors, especially quadcopters, enabled by composite frames, lithium-polymer batteries, and brushless motors\cite{segui2014novel}. Quadcopters combine VTOL and precise hover for monitoring, inspection, and short-range delivery, while hybrid VTOL designs trade hover agility for fixed-wing endurance in longer missions. In the 2010s, swarm robotics expanded UAV capabilities through distributed coordination for coverage and mapping\cite{chung2018survey}. Consensus-style control and potential-field planners support formation keeping, collision avoidance, and task allocation under communication constraints. More recent work targets resilience and safety via fault-tolerant and robust control (e.g., $H_\infty$) to handle turbulence, failures, and payload changes, alongside modular designs that enable rapid payload swapping\cite{lopez2015robust}. Together, these developments improve the scalability and reliability needed for LAWN-scale fleet operations in low-altitude airspace.

\subsubsection{Signal Processing and Aerospace Electronics - Remote Sensing, UAV Detection, and Tracking}

Signal processing and aerospace electronics underpin LAWN sensing, localization, and interference management. The field grew from post-WWII radar, where early UAVs used pulse-Doppler sensing for navigation and target detection. Resolution and clutter limits in complex terrain soon drove richer processing, and satellite remote-sensing ideas later migrated to UAV platforms for environmental monitoring\cite{tang2015drone}. Doppler/time-frequency methods enabled micro-motion analysis, helping distinguish UAV signatures from clutter in urban canyons or forested scenes. Multi-sensor fusion further strengthened tracking and identification by combining radar, RF sensors, and cameras. Kalman and particle filters fuse heterogeneous measurements to reduce position/velocity uncertainty under fog, multipath, and partial occlusion. RF localization techniques such as time-difference-of-arrival (TDoA) and angle-of-arrival (AoA) enable triangulation across distributed receivers for more precise tracking\cite{nie2021uav}. Since the 2010s, AI-enhanced processing has incorporated LiDAR, UWB, and camera data to improve localization in dense urban and indoor environments, while beamforming and spatial filtering mitigate interference by steering beams toward desired links and placing nulls toward interferers. Recent work emphasizes real-time fusion and spectrum sensing to cope with fast-changing LAWN topologies and regulatory requirements (e.g., Remote ID). Integrated sensing and communication (ISAC) has become a key paradigm, reusing shared waveforms for localization, channel estimation, and data transmission, while wavelet features and deep neural networks (DNNs) support RF fingerprinting, UAV identification, and anomaly detection for security and safety\cite{hou2025target}.

\subsubsection{Additional Disciplines and Overall Convergence}
Beyond the information technology–oriented disciplines discussed above, aerospace engineering, materials science, and regulatory science provide essential foundations for LAWNs. Aerospace engineering contributes aerodynamics and structural design optimized for low-altitude flight. Materials science improves endurance and payload capacity through lightweight, durable airframes, reducing weight while increasing strength. Regulatory science shapes scalable airspace integration via policy, risk assessment, and standardization, complementing technical work in communications and control. Together with communications, AI, robotics, and signal processing, these disciplines converge into LAWN as resilient ``intelligent skies'' infrastructure that supports heterogeneous networking with stronger reliability, safety, and security.

\subsection{Contributions \& Scopes}
This paper presents a comprehensive overview of the LAWN, which represents a transformative paradigm integrating aerial drones into intelligent, reconfigurable 3D network architectures operating below $3,000$ m. The primary contributions of this work include a detailed historical analysis of LAWN’s multi-disciplinary evolution, a structured architectural framework, an exploration of AI and signal processing roles, practical case studies, and identification of research challenges. In Section II, we outline LAWN’s architecture and applications, defining altitude-based layers and functional planes, reviewing standardization efforts like 3GPP, and discussing the roles of AI and signal processing in enhancing network performance. Section III focuses on signal processing for LAWN, covering performance metrics, signaling and receiver design, localization and tracking methods, and multi-functionality co-design. Then, Section IV delves into AI for LAWNs, introducing fundamentals, enhanced functionalities, e.g., control, resource allocation, security, and emerging paradigms like generative models and LLMs for mission planning. Section V presents a case study, detailing an integrated signal processing and AI framework. Finally, we identify research challenges and open issues in Section VI and draw our conclusions in Section VII. Our hope is that this paper would provide a holistic understanding of LAWN’s development, functionalities, and future directions, serving as a reference for researchers, engineers, and policymakers shaping the intelligent skies.

\textit{Notations:} Unless otherwise specified, bold lowercase and uppercase letters, e.g., $\mathbf{a}$ and $\mathbf{A}$, denote vectors and matrices, respectively. $\mathbb{C}^{M\times N}$ indicates the $M\times N$-dimensional complex-valued space. $\left(\cdot\right)^{\mathsf{T}}$, $\left(\cdot\right)^{-1}$, and $\left(\cdot\right)^{\mathsf{H}}$ denote the transposition, inverse operation, and Hermitian, respectively. $\|\cdot\|$, $\textrm{tr}\left(\cdot\right)$, and $\mathbb{E}\{\cdot\}$ are respectively the $\ell_{2}$-norm, trace, and expectation of a matrix. $\mathcal{N}\left(\textbf{a},\textbf{B}\right)$ denotes a Gaussian distribution with mean $\textbf{a}$ and covariance matrix $\textbf{B}$, and $\mathbf{I}_{M}$ is the $M\times M$ identity matrix. The function $Q^{-1}(\cdot)$ is the inverse of Gaussian $\mathcal{Q}$ function, i.e., $Q(x)=\frac{1}{\sqrt{2\pi}}\int_{x}^{\infty}\exp(-\frac{t^{2}}{2})\textrm{d}t$. $\mathbf{A}\succeq\mathbf{B}$ and $\mathbf{A}\succ\mathbf{B}$ denotes $\mathbf{A}-\mathbf{B}$ is positive semidefinite and positive definite, respectively.

\section{The LAWN Architectures}
\begin{table*}[!ht]
\caption{Existing overview papers on the LAWN}
\centering
\setlength{\tabcolsep}{3pt}
\renewcommand{\arraystretch}{1.5}\label{survey_LAWN}
\begin{tabular}{|p{0.07\linewidth}|p{0.04\linewidth}|p{0.58\linewidth}|p{0.26\linewidth}|}
\hline
\textbf{Existing Work} & \textbf{Year} & \textbf{Technical aspects} & \textbf{Research discipline} \\
\hline
\cite{Bekmezci2013AdHoc_FANET,Rezwan2021Electronics_FANET_RL} & 2013 & Flying ad-hoc network (FANET) architectures, routing, medium access control, and mobility models & Ad-hoc networking\\
\hline
\cite{Pasandideh2023JFR_FANET_SLR} & 2023 & Protocols for aerial swarms and evaluation metrics & Ad-hoc networking\\
\hline
\cite{kim2019unmanned,fan2020review,coppola2020survey,khalid2023control} & 2020 & UAV swarming and coordination, state estimation, and distributed control & Aerial robotics, navigation, and control \\
\hline
\cite{9462539, zhou2024current} & 2022 & Aerial manipulation, force interaction, multi-modal perception, and design challenges & Aerial robotics, navigation, and control \\
\hline
\cite{chang2023review} & 2023 & Navigation in GNSS–denied environments, visual–inertial odometry, and SLAM & Aerial robotics, navigation, and control \\
\hline
\cite{dimmig2024survey} & 2024 & Aerial robot simulators; physics and sensor fidelity; multi-agent capabilities; benchmarking & Aerial robotics, navigation, and control \\
\hline
\cite{Hamissi2023CSUR_UTM} & 2023 & UTM architecture, services, and standards & Air traffic management and regulation \\
\hline
\cite{Arafat2024Drones_UAMSurvey} & 2024 & UAM networking, safety, spectrum, and latency & Air traffic management and regulation \\
\hline
\cite{aposporis2024review,panov2024critical} & 2024 & Global and regional frameworks for UTM, comparative policy analysis, and data requirements & Air traffic management and regulation \\
\hline
\cite{kanellakis2017survey} & 2017 & Computer vision-based estimation, mapping, tracking, and control & AI for aerial systems \\
\hline
\cite{sai2023comprehensive, cheng2023ai} & 2023 & AI for perception and communications in UAVs & AI for aerial systems \\
\hline
\cite{ekechi2025survey} & 2025 & Multi-agent reinforcement learning for control and coordination of UAVs & AI for aerial systems \\
\hline
\cite{Fotouhi2019COMST_CellularSurvey} & 2019 & Practical deployment, standardization, regulation, and security for cellular-connected UAVs & Cellular network and mobile systems \\
\hline
\cite{Geraci2022COMST_CellularUAV} & 2022 & Radio access network design, interference management, and evolution to 6G & Cellular network and mobile systems \\
\hline
\cite{Ceviz2023arXiv_UAVSecuritySurvey} & 2024 & Detection and mitigation strategies for LAWNs & Cybersecurity for aerial systems \\
\hline
\cite{colomina2014unmanned} & 2014 & Photogrammetry and remote sensing workflows and data processing & Remote sensing and geomatics\\
\hline
\cite{mogili2018review} & 2018 & Precision agriculture using drone systems: crop monitoring, spraying, and workflow integration & Remote sensing and geomatics \\
\hline
\cite{lyu2023unmanned} & 2023 & Search and rescue applications, task assignment, and path planning & Remote sensing and geomatics\\
\hline
\cite{dadrass2024unmanned} & 2024 & Geophysical sensing: magnetometry, gravimetry, electromagnetic, and ground-penetrating radar & Remote sensing and geomatics\\
\hline
\cite{Zeng2019ProcIEEE_UAV_Tutorial, Khawaja2019COMST_A2GSurvey} & 2019 & System-level UAV communications, A2G channel modeling, and trajectory planning & Wireless communications\\
\hline
\cite{Nomikos2022arXiv_MaritimeUAVSurvey} & 2022 & Maritime UAV communications, over-sea propagation, and deployment& Wireless communications\\
\hline
\cite{Zeng2024COMST_CKM_Tutorial} & 2024 & Channel knowledge maps, environment-aware planning, and radio map construction& Wireless communications\\
\hline

\end{tabular}
\end{table*}

Building on the historical convergence of disciplines that have shaped LAWNs, this section translates the ``why LAWNs'' question into ``what a LAWN looks like''. Numerous survey, tutorial, and overview papers have examined related aspects, providing foundational insights into the evolution from ground-based to 3D aerial networks, as summarized in Table \ref{survey_LAWN}. \revise{ Notice that although LAWNs overlap with several existing aerial-networking
paradigms, their scope is broader. For instance, FANETs primarily address distributed
routing and medium access among UAVs, while UAV-assisted cellular
systems treat UAVs as aerial users, relays, or base stations within a
cellular architecture. UTM ecosystems focus on airspace authorization,
traffic coordination, and operational safety, whereas aerial swarm
networks emphasize cooperative motion, formation control, and task
allocation. A LAWN integrates these otherwise separate functions into
a unified low-altitude infrastructure spanning communication, sensing,
computing, control, and airspace management.} In this section, we will introduce the LAWN architecture, followed by standards progress, and the roles of AI and signal processing.

\subsection{Architectural Perspective: Altitude-based Layers and Functional Planes}
The LAWN adopts a dynamically reconfigurable 3D architecture to support high-mobility operations under variable NLoS propagation, fast handovers, and evolving regulatory constraints. A practical way to organize this design is to view it from two complementary angles: (i) \emph{altitude-based layers} that align operations with environment and rules, and (ii) \emph{functional planes} that provide the core capabilities for integrated connectivity, sensing, and control.

\subsubsection{Altitude-based Layers}
\begin{figure*}
    \centering
    \begin{subfigure}{0.5\linewidth}
        \centering
        \includegraphics[width=\linewidth]{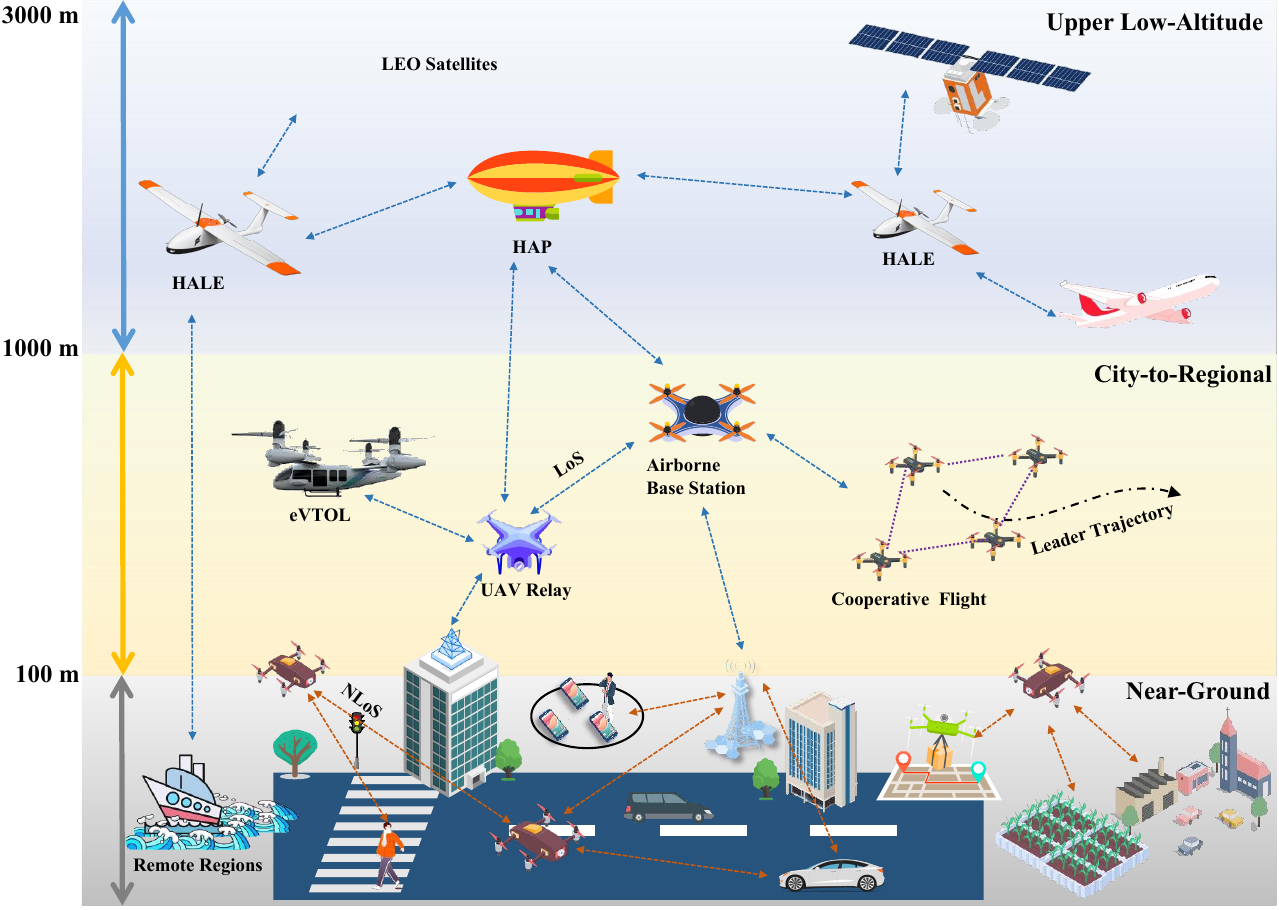}
        \caption{ Altitude-based layers. \label{UPa}} 
    \end{subfigure}
\hfill
    \begin{subfigure}{0.47\linewidth}
        \centering
        \includegraphics[width=\linewidth]{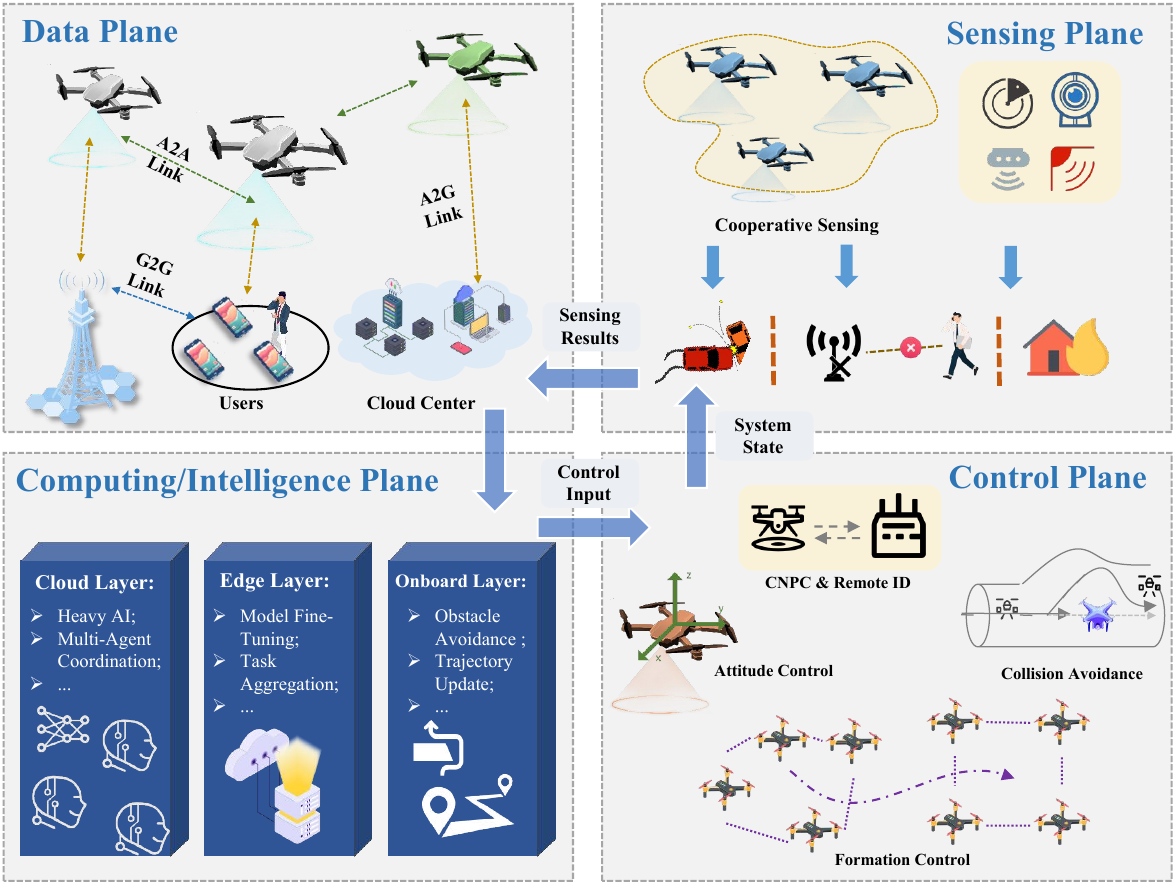}
        \caption{Functional planes.\label{UPb}} 
    \end{subfigure}
\caption{Architectural perspective of LAWNs in terms of altitude-based layers and functional planes. (a) Altitude-based layering partitions low-altitude airspace into near-ground, city-to-regional, and upper low-altitude tiers. (b) Functional planes abstract the end-to-end operation: the data plane transports mission payloads; the control plane enables safety-of-flight functions; the sensing plane fuses observations into situational awareness; and the computing/intelligence plane spans resources to support real-time inference, planning, and multi-agent coordination.}

\end{figure*}
A useful working stratification partitions low-altitude airspace into three tiers: below $100$ m, $100$--$1{,}000$ m, and $1{,}000$--$3{,}000$ m. Each tier features distinct channel conditions, risk profiles, and infrastructure choices.

\paragraph{Below $100\,\mathrm{m}$ (near-ground)}
This tier targets dense urban and cluttered environments, where blockage and NLoS are frequent. Reliable operation typically relies on dense terrestrial support complemented by opportunistic low-flying relays or multi-hop meshes to stabilize C2 and reduce outage-induced risk. Regulations often emphasize tighter operational constraints, including Remote ID and geofencing in populated areas. A {terrestrial-network-centric} architecture with vertical beam/tilt optimization and on-demand relaying is usually the most effective.

\paragraph{$100$--$1{,}000\,\mathrm{m}$ (city-to-regional)}
This tier supports corridor-style mobility, regional surveillance, and eVTOL transport, where LoS is dominant. However, mobility and interference management become key challenges. Fast UAV motion increases handover frequency and expands uplink interference footprints across many cells. Hence, trajectory-aware mobility prediction, AI-assisted beam/cell selection, and interference-aware scheduling are essential. In practice, continuous service is achieved by combining (i) enhanced terrestrial macro sites on tall structures, (ii) UAV relays to patch coverage gaps along aerial corridors, and (iii) airborne base stations that create mission-specific ``capacity/control bubbles.'' BVLOS operations and UTM/U-space integration typically become more relevant in this tier.

\paragraph{$1{,}000$--$3{,}000\,\mathrm{m}$ (upper low-altitude)}
This tier enables wide-area sensing, backhaul, and long-range tasks (e.g., environmental mapping, border monitoring) and begins to interface with HAPs/LEO satellites. LoS is nearly constant, but longer ranges tighten link budgets and shared use with crewed aviation raises requirements on cooperative surveillance and traffic awareness. Connectivity often shifts toward aerial backhaul and satellite support, with high-altitude long-endurance platform considered for persistent coverage and edge resources.

\paragraph{Cross-layer considerations}
To ensure seamless operations of the LAWN across different layers, multi-band and multi-radio access technologies are expected that can adapt to LoS/NLoS conditions and regulatory constraints. In addition, the beam management in the vertical direction should provide robust interference control for aerial users. Moreover, the route planning should co-optimize flight paths with coverage and handover costs and tightly integrate with UTM/U-space services for dynamic corridor assignment.

\subsubsection{Functional Planes}
Complementing the altitude layers, the LAWN adopts four functional planes, i.e., \emph{data}, \emph{control}, \emph{sensing}, and \emph{computing/intelligence}, to modularize capabilities and enable cross-plane collaboration. Conceptually, the planes act as a ``nervous system'': they exchange state, reconfigure resources, and co-optimize safety, efficiency, and mission performance.

\paragraph{Data plane}
The data plane carries mission payloads, e.g., video, mapping information, software updates, and sensory data, across three fundamental link types: A2G, G2A, and air-to-air (A2A). Low-altitude dynamics cause rapid LoS/NLoS switching, Doppler shifts, and intermittent blockage. A scheduler tags bursts with LoS/NLoS state, path loss, and blockage risk, diverting packets via relays or prioritizing time-critical flows when needed. For example, elastic/delay-tolerant (e.g., raw logs) can be buffered first until channel conditions improve. The latency-sensitive flows (e.g., live video and control signals) prefer stable LoS routes for transmission. The group-oriented data (e.g., shared perception maps) can be broadcast across the aerial mesh via A2A links. 

\paragraph{Control plane}
The control plane sustains safety-of-flight communications and airspace compliance, carrying control and non-payload communications (CNPC) traffic, Remote ID, and conformance data with ultra-high reliability and low latency ($<10$ ms, $>99.999\%$ reliability). Primary links favor robust, interference-tolerant waveforms and protected bands, while heterogeneous diversity (e.g., LEO satellite backups) preserves command continuity when corridors lose terrestrial coverage. Packet integrity and time synchronization are enforced to support collision avoidance, geofencing, and formation-keeping. State beacons and concise commands keep swarms coherent without overloading the channel, and control remains logically isolated from payload data to prevent contention under stress.

\paragraph{Sensing plane}
The sensing plane provides network-enabled perception and localization by fusing onboard measurements (cameras, LiDAR, mmWave radar, GNSS/IMU) with RF signals and shared observations from nearby nodes. Continuous updates of LoS/NLoS classification, blockage likelihood, and obstacle states feed into local world models that guide routing, beam choice, and speed/altitude advisories. Cooperative exchange over A2A/A2G links turns narrow fields of view into area-wide situational awareness, while confidence scores and timeliness tags ensure consumers can trade accuracy for latency. When visibility degrades, the plane down-selects to the most reliable modalities and elevates risk signals to the control plane.

\paragraph{Computing/intelligence plane}
The computing/intelligence plane closes the loop by jointly allocating processing and intelligence across onboard, edge, and cloud resources. It handles both conventional workloads and AI-related tasks (inference, planning, learning), subject to constraints on latency, energy, and link capacity. Time-critical functions such as obstacle avoidance, short-horizon trajectory updates, safety monitoring, and link adaptation run locally on the UAVs to remain robust against brief outages or backhaul degradation. By contrast, heavier workloads, including high-resolution perception, multi-agent coordination, model training or fine-tuning, and large-scale map reconstruction, can be offloaded to nearby edge servers or the cloud whenever uplink conditions permit. Techniques such as knowledge distillation and model compression are used to push updated intelligence back to resource-constrained platforms without overwhelming their compute or memory budgets. Under congestion or tight power budgets, it prioritizes essential control loops and minimal sensing and inference pipelines that are needed for safe operation at real-time cadence, while deferring or downscaling non-critical analytics, logging, and background learning tasks. \revise{
For generative AI, this functional split is particularly important
because large models are generally impractical for onboard execution
on small multirotors. The onboard tier therefore retains lightweight
perception, distilled inference, and safety-critical control, whereas
ground edge servers perform latency-sensitive generative inference,
multi-UAV data fusion, and local model adaptation. The cloud supports
computationally intensive model pretraining, large-scale scenario
generation, and global model updates. During connectivity disruptions,
each UAV falls back to its onboard safety policy without relying on
remote generative inference.}

\paragraph{Cross-plane scheduling}
The four planes are designed to interact continuously rather than operate in isolation. The sensing plane maintains up-to-date estimates of LoS/NLoS conditions, blockage likelihoods, and obstacle trajectories. The control plane consumes this information to adjust corridors, speeds, and altitudes, and to revise separation rules or formation geometries when risk increases. The data plane subscribes to the same context to steer routing, choose beams and cells, and decide when to insert relays or switch between A2G and A2A paths. The computing/intelligence plane remains in the loop through a three-tier hierarchy spanning onboard, edge, and cloud. In this way, the four planes collectively provide a coordinated substrate that can adapt in real time to both airspace conditions and mission demands.

\subsection{Standardization Progress}
As the LAWN evolves into a coordinated 3D network supporting applications from logistics to UAM, standards bodies have accelerated efforts to address unique challenges such as high-mobility connectivity, interference in LoS environments, spectrum management, security, and regulatory compliance. Major organizations like 3GPP, ITU-R, IEEE, FAA, EASA, and others have made advancements since 2020, as summarized in Table \ref{tab:lawn-standards}.

\begin{table*}[!ht]
\caption{Standards Progress Relevant to the LAWN}
\label{tab:lawn-standards}
\centering
\setlength{\tabcolsep}{3pt}
\renewcommand{\arraystretch}{1.5}
\begin{tabular}{|p{0.15\linewidth}|p{0.36\linewidth}|p{0.1\linewidth}|p{0.3\linewidth}|}
\hline
\textbf{Standardization Body} & \textbf{Key Focus} & \textbf{Year} & \textbf{Impact on LAWN} \\
\hline
3GPP Release 16 \cite{ericssonR16overview} & 5G NR Phase-2: MIMO/beamforming enhancements & 2020 & Improves aerial-UE robustness and interference handling \\
\hline
3GPP Release 17 \cite{3gppNTN} & First normative NTN features; mmWave enhancements & 2022 & Enables satellite–UAV hybrid architectures and extended-range LAWN backhaul \\
\hline
3GPP Release 18 \cite{3gpp2024service} & AI/ML for trajectory and resource optimization; UAV connectivity enhancements & 2024 & Enables adaptive 3D networks with improved mobility and interference mitigation for UAV integration \\
\hline
ITU-R WRC-23 \cite{itur2024future} & Spectrum allocation (e.g., 5030-5091 MHz) for UAS command-and-control & 2023 & Ensures interference-free C2 links in low-altitude environments, supporting global harmonization \\
\hline
IEEE 802.11bd \cite{ieee2023ieee} & Low-latency A2A communications for high-mobility scenarios & 2023 & Supports swarm operations and V2X-adapted links in urban drone networks\\
\hline
IEEE 802.15.4 \cite{ieee154_2020,ieee154_2024} & Low-power, low-cost, low-data-rate wireless communication over short distances & 2020-2024 & Suits payload/energy-constrained drones and dense local sensing/telemetry \\
\hline
\end{tabular}
\end{table*}

\subsubsection{Cellular Integration for the LAWN}
3GPP has progressively incorporated UAVs into cellular systems by treating drones as aerial user equipment (UE) and, increasingly, as aerial access/relay nodes. Release~15 established baseline aerial connectivity with height-aware interference mitigation and C2-grade URLLC targets (e.g., $\sim$100~kbps and sub-$50$~ms latency) \cite{tr36777}. Releases~16-17 extended MIMO/beam management and aerial channel modeling, and introduced NTN and mmWave options to enhance coverage and backhaul for BVLOS operations \cite{ericssonR16overview,3gppNTN}. Release~18 further emphasized AI/ML-assisted mobility and resource management and incorporated ISAC-oriented directions for joint localization and communications \cite{3gpp2024service,sa5AIMLmgmt}. These capabilities have been supported by field trials demonstrating feasible aerial mobility performance within regulatory altitude limits \cite{qualcomm20245g}. \revise{
It is important, however, to distinguish LAWNs from 3GPP NTNs. As discussed above, LAWNs integrate communication, sensing, computing, and control for highly mobile aerial services below 3,000 m, whereas NTNs provide wide-area access through satellites or high-altitude platforms. NTN procedures primarily address long propagation delays and large but predictable orbital Doppler shifts. By contrast, LAWNs experience abrupt mobility, frequent LoS/NLoS transitions, local blockage, and rapidly varying interference, thereby requiring trajectory-aware handover, agile beam tracking, and adaptive channel estimation. NTN can therefore complement LAWNs through backhaul, coverage extension, and fallback connectivity.}

\subsubsection{ITU-R and Spectrum Allocation}
ITU-R coordinates global spectrum use for UAS communications and coexistence with incumbent aeronautical and terrestrial services. WRC-19 identified candidate bands (including portions of the $5$~GHz range) for UAS C2, and WRC-23 allocated $5030$--$5091$~MHz for safety-critical UAS operations with stringent protection criteria \cite{itur2024future,kamali2018aeromacs}. Looking ahead, WRC-27 study items are expected to address 6G-era LAWN spectrum demands (potentially including bands above $100$~GHz), while continued regional divergence motivates spectrum-sharing and interference-mitigation mechanisms.

\subsubsection{IEEE Standards for LAWNs}
IEEE standards complement 3GPP’s wide-area focus by supporting local, ad-hoc, and short-range connectivity. In particular, IEEE~802.11bd targets low-latency, high-mobility A2A/V2X-style links, while IEEE~802.15.4 provides low-power short-range telemetry suitable for dense sensing and intra-vehicle networking \cite{ieee2023ieee,ieee154_2020,ieee154_2024}.

\subsubsection{Aviation Regulatory Standards}
Aviation regulators worldwide are formalizing Remote~ID, BVLOS approval pathways, and UTM/U-space integration through rulemaking and large-scale trials, as shown in Table~\ref{table1}. Collectively, these programmes translate technical standards from 3GPP, ITU-R, and IEEE into certifiable requirements, helping ensure LAWN deployments remain interoperable, auditable, and publicly acceptable.

\subsection{The Roles of AI and Signal Processing}
To support LAWNs in dynamic 3D environments, signal processing and AI play complementary roles. Signal processing provides the core machinery for reliable communication and sensing—detection, estimation, synchronization, and interference management under aerial propagation conditions. For instance, array processing and beamforming steer energy toward intended links while suppressing dominant interferers in dense swarms. With the rise of ISAC, signal processing further drives co-design of waveforms, pilots, and receivers so that the same physical-layer resources simultaneously deliver data and extract environmental information. AI complements signal processing by enabling predictive, autonomous, and scalable decision-making, turning largely static designs into adaptive policies that operate under energy limits and multi-agent interactions. When tightly coupled with signal processing and deployed hierarchically across onboard, edge, and cloud resources, AI enables LAWNs to maintain reliability and safety while adapting to fast-changing environments and mission demands.

\section{Signal Processing for LAWNs}
Signal processing is the engine that turns LAWN’s architectural ideas into reliable, safe, and efficient operations in dynamic low-altitude airspace. 
In this section, we aim for providing an overview of signal processing applications in LAWNs, covering a broad spectrum of topics while placing particular emphasis on multi-functionality co-design, which integrates multiple functionalities of communications, sensing, and control in resource-constrained aerial settings.

\subsection{Performance Metrics from the Signal Processing Perspective}
Performance metrics serve as critical benchmarks for evaluating if the whole system works well. In what follows, we summarize key metrics categorized by different core functionalities. 

\subsubsection{Communication-Related Metrics}
The throughput or data rate is a primary metric for LAWN communications, quantifying the effective data transmission rate for applications like real-time video relay. 
End-to-end delay is captured by the {latency}, which includes access/scheduling, queuing, propagation, and baseband/stack processing. Signal processing techniques like channel prediction help to reduce the latency. Reliability is commonly assessed via the bit error rate (BER), with safety-critical targets like BER$<10^{-5}$. Channel coding and advanced receiver design are general signal processing tools for improving error performance. When the SINR falls below a certain threshold $\gamma$, the link enters \emph{outage}. The corresponding outage probability $\textrm{Pr}[\textrm{SINR}<\gamma]$ directly relates to link availability.

\subsubsection{Sensing and Localization Metrics}
Signal processing enables robust environmental awareness and precise localization. The estimation accuracy is commonly quantified by the mean square error (MSE) of position/velocity estimates or by the theoretical Cram\'{e}r–Rao Bound (CRB), $\mathrm{CRB} = \left[ \mathbf{I}(\boldsymbol{\theta}) \right]^{-1}$, where $\mathbf{I}(\boldsymbol{\theta})$ is the Fisher information matrix derived from received signal parameters. For UAV/obstacle detection and remote ID enforcement, the detection probability $P_d$ and false alarm rate $P_{fa}$ are two critical metrics. Constant false alarm rate (CFAR) processing typically aims for $P_d > 99\%$ at $P_{fa} < 10^{-3}$ to $10^{-6}$, with Neyman–Pearson or sequential detectors optimizing the trade-off. In addition, spatial and temporal resolution determines the granularity of environmental maps and tracking updates and can be related to waveform ambiguity properties, e.g., main-lobe widths in delay/Doppler/angle.

\subsubsection{Control Metrics}
Control metrics are essential for evaluating the stability, responsiveness, and efficiency of control loops that integrate aerial drones with terrestrial systems. 
A standard metric is the linear–quadratic regulator (LQR) cost that penalizes state deviation and control effort,
\begin{align}
    J = \int_{0}^{\infty} \left(\boldsymbol{\theta}(t)^{\mathsf{T}} \textbf{Q} \boldsymbol{\theta}(t)+\textbf{u}(t)^{\mathsf{T}}\textbf{R}\textbf{u}(t)\right)\textrm{d} t,
\end{align}
where $\textbf{u}(t)$ is the control input, $\textbf{Q}\succeq 0$ and $\textbf{R}\succ 0$ are two matrices that define the trade-off in the LQR optimization. Because control performance depends on timely state updates, the age of information (AoI) measures how communication latency degrades control quality. In addition, for swarms and relays, formation-keeping error, consensus time, and {string stability} are key, alongside graph-theoretic connectivity measures relevant to A2A links.

\subsubsection{Other Performance Metrics}
In addition to the core metrics for communication, sensing, and control, several other performance indicators are crucial for LAWNs, particularly addressing energy constraints, safety risks, scalability, and user-centric quality. Energy efficiency is vital for battery-limited UAVs. 
Risk and safety metrics focus on mitigating hazards in open airspace, with collision probability serving as a key indicator. Scalability metrics evaluate network performance under growing node densities. 
Finally, quality of experience (QoE), security and privacy metrics capture user satisfaction and resilience to threats, including video QoE, confidentiality/integrity guarantees, and privacy leakage risk.

\subsection{Signaling and Receiver Design}
Signaling and receiver design are fundamental to enabling robust connectivity and data exchange in low-altitude environments. Signal design in LAWNs must accommodate highly dynamic channels and multi-functional requirements. Unlike static terrestrial links, UAV-based links face rapid Doppler shifts and multipath effects. The channel model can typically be expressed as 
\begin{align}
     h(\tau, t) = \sum_{l=0}^{L-1} \alpha_l(t) \delta(\tau - \tau_l) e^{j 2\pi f_{D,l} t},
\end{align}
where $\alpha_l$ denotes the path gain, $\tau_l$ the delay, $f_{D,l}$ the Doppler frequency, and $L$ is the number of resolvable paths. 
Note that the Doppler shift significantly degrades the communication performance. To enhance signaling robustness in high-Doppler environments, various researches also focus on new waveform designs. Orthogonal time frequency space (OTFS) modulation, a 2D scheme operating in the delay-Doppler (DD) domain, multiplexes symbols on a DD grid and transforms them to the time-frequency domain via the inverse symplectic finite Fourier transform, expressed as 
\begin{align}
X[n, m] = \frac{1}{\sqrt{MN}} \sum_{k=0}^{N-1} \sum_{l=0}^{M-1} x[k, l] e^{j 2\pi (nk/N - ml/M)},
\end{align}
where $x[k, l]$ are DD symbols. OTFS outperforms classic orthogonal frequency division multiplexing (OFDM) in high-mobility scenarios by representing the doubly selective channel in the DD domain, achieving full diversity and lower BER at high Doppler shifts \cite{wei2021orthogonal,yuan2023new}. Similarly, affine frequency division multiplexing (AFDM), a chirp-based waveform, uses the discrete affine Fourier transform to map symbols, given by 
\begin{align}
p_{AT,m} = \frac{1}{\sqrt{M_0}} e^{-j 2\pi c_2 m^2} \sum_{n=0}^{N_0-1} e^{-j 2\pi \left( \frac{1}{M_0} m n + c_1 n^2 \right)} s_n,
\end{align}
where $c_1, c_2$ are chirp parameters\cite{bemani2021afdm}. Choice among OFDM, OTFS, AFDM, or hybrid modulation schemes should be guided by mobility profiles, bandwidth, and multi-function constraints.

\begin{table*}[!ht]
\caption{Factor-Graph–Based Message Passing Algorithms}
\label{tab:fg-mp-survey}
\centering
\setlength{\tabcolsep}{3pt}
\renewcommand{\arraystretch}{1.5}
\begin{tabular}{|p{0.15\linewidth}|p{0.35\linewidth}|p{0.21\linewidth}|p{0.21\linewidth}|}
\hline
\textbf{Algorithm} & \textbf{Technical Aspects} & \textbf{Pros} &\textbf{Cons}\\
\hline
Sum–Product Algorithm (SPA)\cite{kschischang2002factor}&
Belief propagation (BP) with sum–product messages& Near-optimal performance on sparse graphs & Damping needed for convergence and complexity grows with degree and constellation size\\
\hline
Gaussian BP (GaBP) \cite{liu2018gaussian}&
BP with Gaussian messages to solve linear systems &
Scales on sparse systems & Diverge if ill-conditioned\\
\hline
Expectation Propagation (EP) \cite{cespedes2014expectation}&
Project non-Gaussian factors to Gaussians via moment matching&
Stable under strong nonlinearities & Fixed-point tuning required\\
\hline
Generalized Approximate Message Passing (GAMP) \cite{al2017gamp}&
AMP with separable output channels and Onsager correction &
Low per-iteration cost & Sensitive to non-i.i.d. matrices and damping often needed\\
\hline
Orthogonal/Unitary AMP \cite{rangan2019vector,yuan2021iterative}&
AMP variant for right-orthogonally invariant matrices & predictable convergence & SVD/pre-whitening overhead and strong model assumptions\\
\hline
Hybrid Message Passing \cite{li2021hybrid}&
SPA on dominant path + GaBP on less-interfered symbols &
Reduced complexity with favourable performance & Path identification needed\\
\hline
Hybrid Detection \cite{10845213}&
Gaussian mixture model for symbol representation and adaptive during iterations &
Flexibility in controlling error performance and complexity &
Optimal number of Gaussian components is unknown\\
\hline
\end{tabular}
\end{table*}

With the adoption of multi-antenna system, array-based beamforming is a key tool for improving SINR, extending coverage, and mitigating interference in cluttered low-altitude airspace. Given a multiple-input multiple-output (MIMO) channel matrix $\mathbf{H}[k]\in\mathbb{C}^{N_r\times N_t}$, the maximum-ratio transmission/reception for a single stream aligns $\mathbf{w}_t$ and $\mathbf{w}_r$ with the dominant singular vectors of $\mathbf{H}[k]$, while interference-aware designs trade beam gain for null placement. In LAWNs, the desired beam depends on instantaneous 3D geometry such as altitude, bearing, and their predicted evolution. Therefore, geometry-aided tracking techniques can stabilize beams and reduce access/handover latency.

In the context of ISAC, signal waveforms are co-optimized for dual functionality. Let $s(t)$ denote the baseband transmit signal, with ambiguity function
\begin{align}
A(\tau, f_d) = \int s(t) s^*(t - \tau) e^{j 2\pi f_d t} dt.
\end{align}
Communication favors high mutual information or SINR, while sensing favors narrow mainlobes and low sidelobes of $|A(\tau, f_d)|$. A generic ISAC design poses a multi-objective program, i.e.,
\begin{align}\label{ISAC_O1}
\max_{s(t)}& \ \lambda_1\,\textrm{SINR} - \lambda_2\,\mathrm{SL}(s),
\\&\text{s.t.}\ \ \mathbb{E}\{|s(t)|^2\}\!\le P
\end{align}
where $\mathrm{SL}(s)$ denotes the sidelobe level (integrated sidelobe level or peak sidelobe level) achieved by using $s(t)$, $\lambda_1$ and $\lambda_2$ denote parameters for balancing sensing/communication performance. Alternatively, a communication-centric or sensing-centric signaling design can be considered. Taking the sensing-centric case as an example, we aim to minimize the CRB of parameter $\bm{\theta}$ while maintaining the communication rate, i.e.,
\begin{align}\label{ISAC_O2}
\min_{s(t)}& \ \mathrm{tr}\!\big\{\mathbf{I}^{-1}(\bm{\theta})\big\}
\\& \text{s.t.}\ \ R(s)\!\ge\!R_{\min},
\end{align}
where $R(s)$ is the achievable rate. Both problems \eqref{ISAC_O1} and \eqref{ISAC_O2} involve functional optimization. In practice, $s(t)$ is drawn from structured families, so the task can be reduced to sequence, beamforming, or filtering designs.

\begin{figure}[!t]
\centering
\includegraphics[width=.5\textwidth]{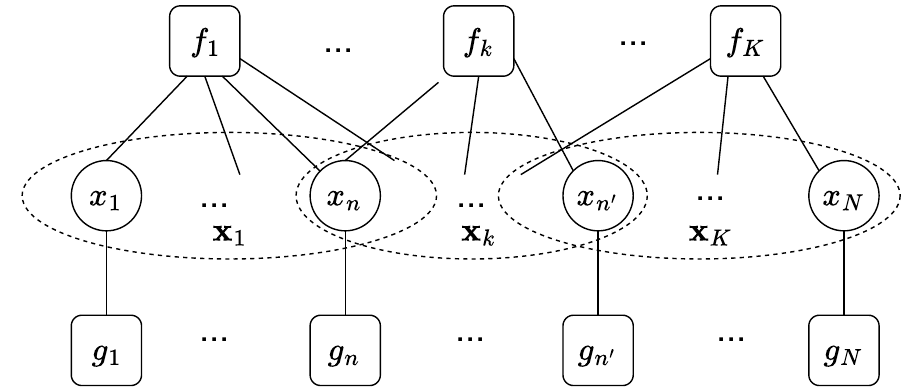}
\caption{{A general factor graph representation for the factorization of $p(\mathbf{y}|\mathbf{x}) $. The shorthand notations $g_n = p(x_n)$ and $f_k  = p(y_k|\mathbf{x}_k)$.}}%
\label{FG}%
\end{figure}

Receiver design in LAWNs prioritizes signal processing to decode signals under channel fadings. Consider a general input-output relationship with the standard linear model $\mathbf{y} = \mathbf{H} \mathbf{x} + \mathbf{n}$,
where $\mathbf{y}\in \mathcal{C}^{K}$, $\mathbf{H}^{K\times N}$, and $\mathbf{x}\in \mathcal{C}^{N}$ denote the receive vector, channel matrix, and transmit symbol vector, respectively. Similar to channel estimation, a linear MMSE (LMMSE) receiver can be adopted by choosing an equalization matrix $\mathbf{W}$ that minimizes the MSE $\mathbb{E}\{\|\mathbf{x}-\mathbf{W}\mathbf{y}\|^2\}$. The detected symbols are given by $\hat{\mathbf{x}} = \mathbf{W}\mathbf{y}$. LMMSE detector offers a favorable BER performance and complexity with the assumption of Gaussian-distributed data symbols. 
For higher reliability, especially for safety-critical control packets, nonlinear detection such as maximum \emph{a posteriori} (MAP) can be applied by maximizing the marginal \emph{a posteriori} distribution of a data symbol $x_n$, i.e., $\hat{x}_n = \arg\max_{x_n\in\mathcal{A}} p(x_n|\mathbf{y})$,
where $\mathcal{A}$ denotes the constellation set to which ${x}_n$ belongs. The MAP detection is performed in a symbol-wise manner and hinges on the knowledge of $p(x_n|\mathbf{y})=\sum_{\mathbf{x}\sim x_n}  p(\mathbf{x}|\mathbf{y})$, 
where $\mathbf{x}\sim x_n$ denotes all elements in $\mathbf{x}$ except $x_n$. The above marginalization operation has a complexity of $\mathcal{O}\left(N\cdot|\mathcal{A}|^{N-1}\right)$, which is prohibitively high. To this end, leveraging message passing on factor graphs can approach near-optimal performance at manageable complexity. A generic factor graph represents the joint distribution of data symbols as a product of local factors, as illustrated in Fig. \ref{FG}:
\begin{align}
    p(\mathbf{x}|\mathbf{y}) \propto \prod_n p(x_n) \prod_k p(y_k|\mathbf{x}_k),
\end{align}
where $\mathbf{x}_k$ denotes all data symbols contributed to the $k$-th received symbol $y_k$, $p(x_n)$ is the \emph{a priori} distribution.
Relying on the factor graph model, message passing algorithms can be implemented to efficiently obtain the marginal distribution. In Table \ref{tab:fg-mp-survey}, we summarize several representative message passing algorithms that have various core ideas and pros/cons. 



\subsection{Localization, Sensing, and Tracking Methods}
Localization, sensing, and tracking are pivotal signal processing tasks in LAWNs. Building on the functional planes introduced earlier, the sensing plane fuses onboard measurements with RF observables to enable precise positioning and environmental awareness.

Let $\mathbf{x}_t$ denote a kinematic state at time instant $t$, e.g., position, velocity, heading and a sensor $i$ will produce the measurement $\mathbf{z}_{i,t} = \mathbf{h}_i(\mathbf{x}_t) \,+\, \mathbf{v}_{i,t}$, where $\mathbf{h}_i(\cdot)$ is the measurement function and $ \mathbf{v}{i,t}$ is the measurement noise\footnote{The measurements can be ToA/TDoA/AoA, carrier phase, received signal strength, or visual/point-cloud projections.}. The goal of localization is to infer the parameter $\mathbf{x}_t$ from the collected measurements $\mathbf{z}_{i,t},~\forall i$. Mathematically, this is solved via a generic nonlinear least squares (LS) minimization, i.e.,
\begin{align}
\hat{\mathbf{x}}_t= \arg\min_{\mathbf{x}_t} \sum_i \| \mathbf{z}_{i,t} - \mathbf{h}_i(\mathbf{x}_t) \|^2_{\mathbf{R}_{i,t}^{-1}},
\end{align}

As for AoA measurement, the measurement function of a 2D position $\mathbf{x}_t = (x_t, y_t)$ and the $i$-th sensor at $\mathbf{x}_{i,t} = (x_{i,t}, y_{i,t})$ can be expressed as $\mathbf{h}_i(\mathbf{p}) = \theta_{i,t} = \arctan\frac{x_t-x_{i,t}}{y_t-y_{i,t}}$. By stacking equations for multiple sensors, a linear system can be formed and solved in LS sense. Similarly, ToA obtains range measurement $c \tau_{i,t} = \|\mathbf{x}_t - \mathbf{x}_{i,t}\|$ from time delays $\tau_{i,t}$ at multiple sensors, with $c$ as the speed of light. In multi-UAV systems, cooperative methods can leverage inter-UAV measurements, e.g., relative ranging to enhance accuracy, especially in GNSS-denied low-altitude environments \cite{yuan2024wireless}. 

To maintain continuous estimates of states of drones or objects of interest, tracking extends localization over time for swarm coordination and threat detection. The state evolution of $\mathbf{x}_t$ is usually modelled by $\mathbf{x}_{t} = \mathbf{g}(\mathbf{x}_{t-1},\mathbf{u}_{t-1}) + \mathbf{w}_{t-1}$, where $\mathbf{w}_{t-1}$ is the transition noise and $\mathbf{g}(\mathbf{x}_{t-1},\mathbf{u}_{t-1})$ denotes the transition function related to the motion patterns, e.g., constant-velocity, nearly-constant-acceleration, and coordinated turn models.

When both the measurement function and transition function can be written in a linear form, the Kalman filter (KF) is optimal. With $\mathbf{F}_{t}$ and $\mathbf{H}_t$ the linear transition and measurement matrices, the standard recursion reads
\begin{align}\label{kf_predict}
\textbf{Predict:}\quad &\hat{\mathbf{x}}_{t|t-1} = \mathbf{F}_{t-1}\hat{\mathbf{x}}_{t-1|t-1} + \mathbf{B}_{t-1}\mathbf{u}_{t-1}, \\
&\mathbf{P}_{t|t-1} = \mathbf{F}_{t-1}\mathbf{P}_{t-1|t-1}\mathbf{F}_{t-1}^\mathsf{T} + \mathbf{Q}_{t-1}; \\
\textbf{Update:}\quad &\mathbf{K}_t = \mathbf{P}_{t|t-1}\mathbf{H}_t^\mathsf{T}\big(\mathbf{H}_t\mathbf{P}_{t|t-1}\mathbf{H}_t^\mathsf{T}+\mathbf{R}_t\big)^{-1}, \\
&\hat{\mathbf{x}}_{t|t} = \hat{\mathbf{x}}_{t|t-1} + \mathbf{K}_t\big(\mathbf{z}_t - \mathbf{H}_t\hat{\mathbf{x}}_{t|t-1}\big), \\
&\mathbf{P}_{t|t} = \big(\mathbf{I}-\mathbf{K}_t\mathbf{H}_t\big)\mathbf{P}_{t|t-1}.\label{kf_estimate}
\end{align}
Nonlinear measurements or kinematics can be handled by the extended KF (EKF), which linearizes $(\mathbf{f},\mathbf{h})$ via Jacobians $\mathbf{F}_{t-1}=\partial \mathbf{f}/\partial \mathbf{x}$, $\mathbf{H}_{t}=\partial \mathbf{h}/\partial \mathbf{x}$ evaluated at the current estimate, or by the {unscented} KF (UKF), which propagates a set of sigma-points to capture nonlinearity without explicit Jacobians. As for the measurements from multi-modal sensors, particle filtering (PF) can be used, which approximates the \emph{a posteriori} distribution of $\mathbf{x}_t$ with $N_p$ weighted samples $\{\mathbf{x}_t^{(i)},w_t^{(i)}\}_{i=1}^{N_p}$ and resamples to focus on likely states \cite{djuric2003particle}.

In LAWNs, on-board sensors or terrestrial nodes often acquire multiple measurements simultaneously, corresponding to various targets, clutter, or environmental artifacts. Multi-target tracking (MTT) is essential for estimating the states of multiple dynamic objects, such as other UAVs, obstacles, or ground entities, while resolving data association ambiguities, determining which measurement originates from which target amid clutter and missed-detections. The measurement and association model forms the basis for MTT. At time $t$, let $\mathcal{L}_t = \{1, \dots, L_t\}$ index $L_t$ existing tracks and $\mathcal{M}_t = \{1, \dots, M_t\}$ index $M_t$ measurements. Under point-target assumptions, a likelihood matrix captures associations between the $j$-th measurement and $i$-th track, i.e., $\ell_{ij} = p(\mathbf{z}_t^{(j)} \mid \mathbf{x}_{t|t-1}^{(i)})$ for target-generated measurements and $\ell_{0j} = \lambda_c \, c(\mathbf{z}_k^{(j)})$ for clutter, where $\lambda_c$ is the clutter rate and $c(\cdot)$ is the spatial probability density function. The joint probabilistic data association (JPDA) filter computes marginal association probabilities $\beta_{ij}$ for measurement $j$ to track $i$ by summing over feasible association events $\mathbb{A}$, given by
\begin{align}
    \beta_{ij} = \sum_{\mathbf{a} \in \mathbb{A}} \mathbb{I}\{a_i = j\} \, p(\mathbf{a} \mid \mathbf{Z}_t),
\end{align}
 with $\beta_{i0} = 1 - \sum_{j \in \mathcal{M}_t} \beta_{ij}$ for missed detections. 

Multiple-hypothesis tracking (MHT) extends this by maintaining a tree of global hypotheses over time, where each branch represents a sequence of associations. Hypothesis scores, typically LLRs, guide pruning and N-scan backtracking to manage combinatorial growth \cite{blackman2004multiple}. 
In some use cases, when the number of targets, clutter, or environmental artifacts are not pre-known or varying with time, we can resort to random finite set (RFS) filters. By treating tracking as a set estimation problem, RFS methods circumvent the need for explicit data associations, instead modeling targets as unlabeled or labeled random sets\cite{vo2008bayesian}. 

\subsection{Multi-functionality Co-Design} 
\begin{figure*}
    \centering
    \includegraphics[width=\linewidth]{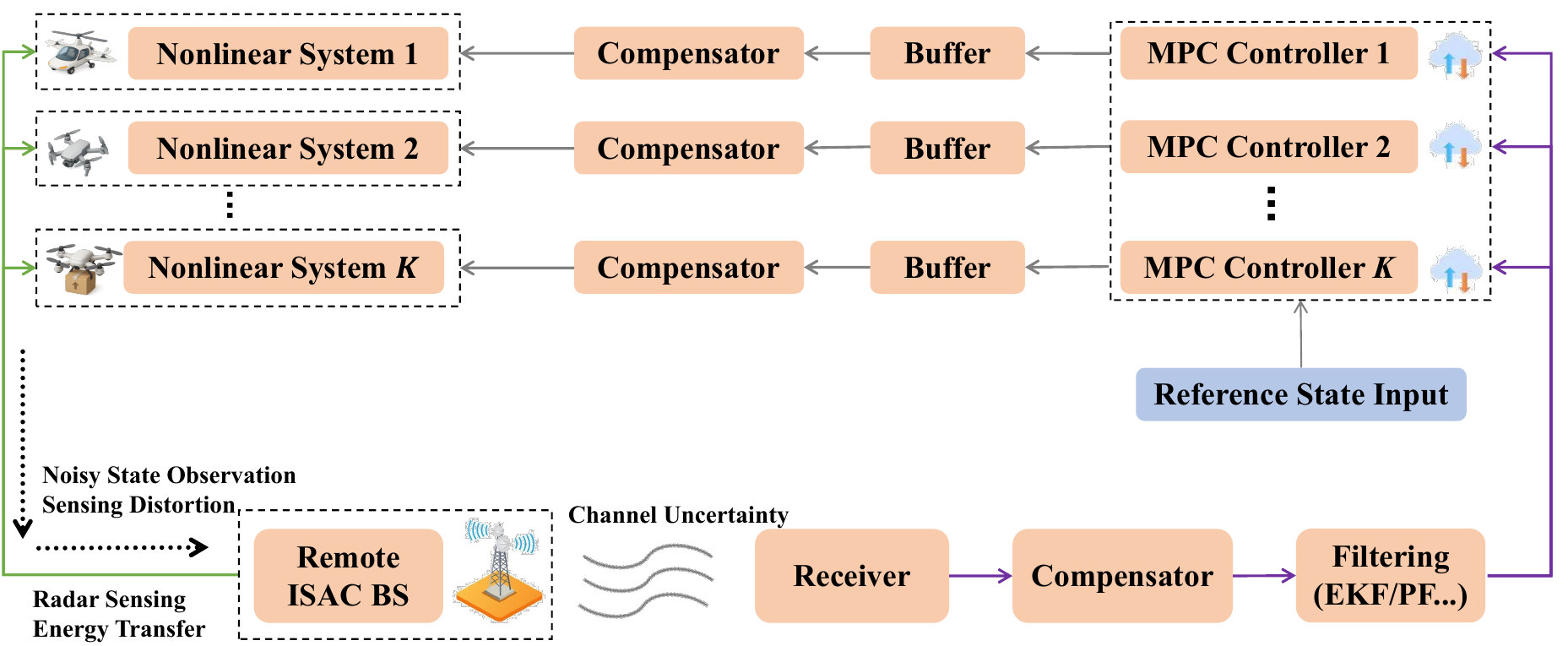}
\caption{Schematic of a representative multi-functionality co-design loop. A remote ISAC base station simultaneously supports communication, radar sensing, and energy transfer, while delivering noisy state observations over an uncertain channel. The receiver-side processing produces state estimates that feed multiple MPC controllers. Control commands are then delivered through buffers/compensators to the corresponding nonlinear plants, capturing practical effects such as communication-induced delay/distortion and actuation/interface dynamics.}

    \label{fig:placeholderqqq}
\end{figure*}
Based on the aforementioned performance metrics, we now turn to multi-functional LAWN designs that explicitly couple communication, sensing, and control. The ultimate objective of intelligent systems deployed in LAWNs is to generate precise control actions through timely and accurate interaction with a dynamic environment. In this context, control is not a simple extension of sensing and communication, but the core mechanism that closes the loop between perception and actuation. Therefore, the transmit strategy and the control inputs should be designed jointly to meet the various desired requirements.

For ease of exposition, we consider a LAWN system where a ground base station (BS) equipped with \(N_t\) antennas serves \(K\) UAVs via ISAC waveforms, as depicted in Fig. \ref{fig:placeholderqqq}. Let \(s_{k,t}\) denote the communication symbol intended for UAV \(k\) at time \(t\), and let \(\mathbf{s}_{\mathrm{d},t} \in \mathbb{C}^{N_t\times 1}\) denote a dedicated radar probing signal. The data symbols \(s_{k,t}\) are assumed mutually independent and circularly symmetric complex Gaussian (CSCG) with zero mean and unit variance, i.e., \(s_{k,t}\sim\mathcal{CN}(0,1)\). The sensing signal satisfies
\(\mathbb{E}\{\mathbf{s}_{\mathrm{d},t}\mathbf{s}_{\mathrm{d},t}^\mathsf{H}\}
=\mathbf{C}_{\mathrm{d},t}\succeq\mathbf{0}\).
The ISAC transmit signal at the BS at time $t$ is then written as
\begin{equation}
    \mathbf{s}_t
    = \sum_{k=1}^{K}\mathbf{w}_{k,t}s_{k,t}
      + \mathbf{s}_{\mathrm{d},t},
    \label{eq:isac-tx}
\end{equation}
where \(\mathbf{w}_{k,t}\in\mathbb{C}^{N_t\times 1}\) is the transmit beamforming vector for UAV \(k\). The corresponding transmit covariance matrix is
\begin{equation}
    \mathbf{C}_t
    \triangleq \mathbb{E}\{\mathbf{s}_t\mathbf{s}_t^\mathsf{H}\}
    = \sum_{k=1}^{K}\mathbf{w}_{k,t}\mathbf{w}_{k,t}^\mathsf{H}
      + \mathbf{C}_{\mathrm{d},t}.
    \label{eq:tx-cov}
\end{equation}
The downlink channel from the BS to UAV \(k\) at instant \(t\) is modeled as a Rician fading channel with both LoS and NLoS components:
\begin{equation}
    \mathbf{h}_{k,t}
    = \sqrt{\rho_0 d_{k,t}^{-2}}\left(
        \sqrt{\frac{\kappa_u}{\kappa_u+1}}\,\mathbf{h}^{\mathrm{LoS}}_{k,t}
        + \sqrt{\frac{1}{\kappa_u+1}}\,\mathbf{h}^{\mathrm{NLoS}}_{k,t}
      \right),
    \label{eq:rician}
\end{equation}
where \(d_{k,t}\) is the BS–UAV distance, \(\rho_0 = G_0\left(\gamma_{\mathrm{s}}/4\pi\right)^2\) with \(G_0\) denotes the reference power gain, \(\gamma_{\mathrm{s}}\) is the carrier wavelength, and \(\kappa_u\) is the Rician factor. The vector
\(\mathbf{h}^{\mathrm{LoS}}_{k,t}\in\mathbb{C}^{N_t\times 1}\) is the deterministic LoS component with unit-modulus entries, while
\(\mathbf{h}^{\mathrm{NLoS}}_{k,t}\in\mathbb{C}^{N_t\times 1}\) represents the scattered NLoS component with i.i.d. entries
\(\mathcal{CN}(0,1)\).
Then, the received signal at UAV  \(k\) is
\begin{align}
    r_{k,t}
   = \mathbf{h}_{k,t}^\mathsf{H}\mathbf{w}_{k,t}s_{k,t}
     + \sum_{\ell\neq k}\mathbf{h}_{k,t}^\mathsf{H}\mathbf{w}_{\ell,t}s_{\ell,t}
     + \mathbf{h}_{k,t}^\mathsf{H}\mathbf{s}_{\mathrm{d},t}
     + z_{k,t},
    \label{eq:rx}
\end{align}
where \(z_{k,t}\sim\mathcal{CN}(0,\sigma_k^{2})\) is the receiver noise. The instantaneous SINR at UAV \(k\) is
\begin{equation}
    \mathrm{SINR}_{k,t}
    = \frac{\bigl|\mathbf{h}_{k,t}^\mathsf{H}\mathbf{w}_{k,t}\bigr|^{2}}
           {\sum\limits_{i\in\mathcal{K}\setminus k}
              \bigl|\mathbf{h}_{k,t}^\mathsf{H}\mathbf{w}_{i,t}\bigr|^{2}
            + \mathbf{h}_{k,t}^\mathsf{H}\mathbf{C}_{\mathrm{d},t}\mathbf{h}_{k,t}
            + \sigma_k^{2}},
    \label{eq:sinr}
\end{equation}
The BS forwards sensing observations of the target region to a UAV controller, which computes the control commands. Without loss of generality, the controlled plant associated with formation \(k\) is modeled as a linear time-invariant discrete-time system: $ \mathbf{x}_{k,t+1} = \mathbf{A}_k \mathbf{x}_{k,t}+ \mathbf{B}_k \mathbf{u}_{k,t}+ \mathbf{v}_{k,t}$, where \(\mathbf{u}_{k,t}\) is the control input. 
The matrices \(\mathbf{A}_k\) and \(\mathbf{B}_k\) are the state transition and input matrices, respectively. The corresponding measurement equation is $\mathbf{y}_{k,t}= \mathbf{G}_k \mathbf{x}_{k,t} + \mathbf{w}_{k,t}$, where \(\mathbf{G}_k\) is the observation matrix.

To facilitate the subsequent system design, let
$\boldsymbol{\chi}
\triangleq
\big\{\mathbf{w}_{k,t},\,\mathbf{C}_{\mathrm{d},t},\,\mathbf{u}_{k,t}\big\}$
be the optimization variables. Then, the vast majority of existing multi-functional design problems can be expressed in the following
compact unified formulation:
\begin{align}
    \min_{\boldsymbol{\chi}}\quad
    & F(\boldsymbol{\chi})
    \label{eq:unity}\\
    \text{s.t.}\quad
    & \mathrm{C1:} \ g_i(\boldsymbol{\chi}) \ge D_i,\quad i=1,\ldots,I, \nonumber\\
    &\mathrm{C2:} \ \mathbf{x}_{k,t+1}
    = \mathbf{A}_k \mathbf{x}_{k,t}
      + \mathbf{B}_k \mathbf{u}_{k,t}
      + \mathbf{v}_{k,t},\nonumber
\end{align}
where $F (\boldsymbol{\chi})$ and \(g_i(\boldsymbol{\chi})\) are generic objective functions and constraints depending on the specific sensing-communication-control-energy requirements. The terms $I$ and $ D_i$ denote the total number of constraints and the $i$th threshold in C1. Constraint C2 enforces the state evolution of each controlled formation according to the underlying plant dynamics. In general, the optimization problem \eqref{eq:unity} is challenging to solve due to the following reasons. First, the optimization problem is non-convex/NP-hard due to the unit modulus constraint of the transmit beamforming vector $\mathbf{w}_{k,t}$. Furthermore, the strong coupling in $\boldsymbol{\chi}$ further hinders achieving a globally optimal solution accounting for the sensing-communication-control performance simultaneously. To compromise, most of the existing works aim at finding highly efficient and locally optimal solutions with low computational complexity. 

It is noted that, if the control inputs are fixed, the problem reduces to a conventional dual-functional ISAC beamforming design, {which has been extensively investigated in the literature\cite{Wu2025JSAC,isac2}}. Motivated by this observation, alternating optimization (AO) is commonly adopted to decouple the design variables. Because the resulting rate/SINR expressions are typically highly nonconvex, advanced tools such as the weighted MMSE (WMMSE) method \cite{WMMSE} and fractional programming (FP) \cite{FP} are often employed to reformulate the original intractable problem into a more tractable approximate form. The transmit beamformers at the BS can then be obtained using standard beamforming techniques. Therefore,  we concentrate on the control-oriented optimization design in what follows by presenting several representative examples.

\subsubsection{Joint Multiple Access and Control Stabilizability Design}

End-to-end delay is captured by the {latency}, which includes access/scheduling, queuing, propagation, and baseband/stack processing. Signal processing techniques like channel prediction help to reduce the latency. Reliability is commonly assessed via the bit error rate (BER), with safety-critical targets like BER$<10^{-5}$. Channel coding and advanced receiver design are general signal processing tools for improving error performance. When the SINR falls below a certain threshold $\gamma$, the link enters \emph{outage}. The corresponding outage probability $\textrm{Pr}[\textrm{SINR}<\gamma]$ directly relates to link availability.
A natural starting point for multi-functional LAWN design is the notion of control stabilizability. Among the various performance indicators, stabilizability is the most fundamental: if the downlink cannot sustain a sufficiently informative flow to each UAV controller, neither sophisticated sensing nor advanced planning can prevent the closed loop. In a LAWN, multiple UAVs with heterogeneous dynamics must share limited time–frequency–space resources. As a result, the BS-side multiple access mechanism directly governs which control loops can be made stabilizable and the stability margin that can be ensured for each UAV.

For the linear system associated with UAV \(k\), classical information-theoretic results relate stabilizability to a minimum required information rate. Let \(\lambda_j(\mathbf{A}_k)\) denote the eigenvalues of the state transition matrix \(\mathbf{A}_k\). The quantity
\begin{equation}
    R_k^{\mathrm{stab}}
    \triangleq
    \sum_{|\lambda_j(\mathbf{A}_k)|>1}
    \log_2\big|\lambda_j(\mathbf{A}_k)\big|
    \label{eq:Rk_stab}
\end{equation}
characterizes the rate threshold that the feedback channel must exceed in order for UAV \(k\) to be mean-square stabilizable over a communication link \cite{tatikonda2004control}. UAVs with more unstable eigenvalues (higher \(|\lambda_j|\)) require a higher \(R_k^{\mathrm{stab}}\), and thus impose more strict demands on the communication and multiple access protocols.

Within this context, rate-splitting multiple access (RSMA) is particularly appealing because its signal structure matches the intrinsic structure of control traffic in LAWNs \cite{mao2022rate}. At each time instant, the BS must deliver two qualitatively different types of information: (i) \emph{UAV-specific} observation and control data, such as state estimates, control increments, or fault indicators; and (ii) \emph{system-level} operational information, such as mission phase, airspace constraints, safety margins, or global coordination commands that are relevant to all UAVs. An RSMA scheme naturally separates these roles by encoding the plant-specific observation/control data of UAV \(k\) into a private stream, while aggregating system-level operational parameters into a single common stream that is decoded by all UAVs.

Under this design, each UAV first decodes the common stream to obtain the shared mission and safety context, and then decodes its private stream to recover the high-rate, plant-specific control updates. Let \(R^{\mathrm{c}}_{k,t}\) and \(R^{\mathrm{p}}_{k,t}\) denote the instantaneous common and private rates experienced by UAV \(k\) at time $t$. Since the common stream must be decodable at all UAVs, its multicast rate is limited by the weakest link,
\begin{equation}
    R^{\mathrm{c}}_{t}
    = \min_{k} R^{\mathrm{c}}_{k,t}.
\end{equation}
Allocating a portion \(C_{k,t}\) of this common-rate budget to UAV \(k\) with \(\sum_{k} C_{k,t} \le R^{\mathrm{c}}_{t}\), the effective rate available to stabilize UAV \(k\) can be written as
\begin{equation}
    R_{k,t}
    = C_{k,t} + R^{\mathrm{p}}_{k,t}.
\end{equation}

From a stabilizability perspective, RSMA provides precisely the flexibility that conventional orthogonal or purely non-orthogonal schemes lack. The common part \(\{C_{k,t}\}\) carries low-rate but globally relevant control information, while the private rates \(\{R^{\mathrm{p}}_{k,t}\}\) supply the high-rate links needed by more unstable UAVs to satisfy \(R_{k,t}\gtrsim R_k^{\mathrm{stab}}\). Moreover, by partially decoding interference, RSMA is inherently more resilient to CSI imperfections, yielding  effective rates and thus more reliable stabilizability margins. In this way, RSMA does not merely improve spectral efficiency. It reshapes the rate region perceived by the controllers and enlarges the set of UAVs that can meet the stabilizability condition in \eqref{eq:Rk_stab}, thereby establishing a control-friendly multiple-access basis for the subsequent multi-functional co-design.

\subsubsection{Rate–Cost Trade-off}
Once stabilizability is ensured, a central question for LAWN-based wireless control is how the available communication rate constraints the achievable control performance\cite{kostina2019rate}. In the considered linear system, the control quality of UAV $k$ over a time horizon $t$ is measured by the LQR cost
\begin{align}
   J_{k,t}
    \triangleq \mathbb{E}\Bigg[
        \sum_{i=0}^{t-1}
        \bigl(
            \mathbf{x}_{k,i}^{\mathsf{T}}\mathbf{Q}_k \mathbf{x}_{k,i}
          + \mathbf{u}_{k,i}^{\mathsf{T}}\mathbf{R}_k \mathbf{u}_{k,i}
        \bigr)
        + \mathbf{x}_{k,t}^{\mathsf{T}}\mathbf{Q}_{1,k} \mathbf{x}_{k,t}
    \Bigg],
    \label{eq:lqr}
\end{align}
where $\mathbf{Q}_k,\mathbf{Q}_{1,k}\succeq\mathbf{0}$ are symmetric positive semidefinite weighting matrices penalizing state deviations, and $\mathbf{R}_k\succeq\mathbf{0}$ is a symmetric positive semidefinite weighting matrix penalizing the control effort in the LQR criterion. Under the classic assumption of essentially perfect and instantaneous state feedback, the design problem reduces to specifying $\mathbf{Q}_k$ and $\mathbf{R}_k$ and synthesizing a stabilizing controller that drives the infinite-horizon LQR cost $J_{k,t}$ as low as possible. 

In LAWN systems, however, the feedback loop for UAV $k$ is closed over a shared wireless medium. The state information must be sensed, encoded, conveyed through an ISAC waveform, and decoded at the controller under stringent spectral and power budgets. As a consequence, if the controller cannot acquire sufficiently informative state estimates, it may no longer be able to construct control inputs that minimize $J_{k,t}$. This motivates a fundamental question, i.e., for a prescribed control performance level $\bar{J}_k$, what is the minimum information rate required to sustain it? Some recent studies have investigated the corresponding characterization of the \emph{rate-control cost trade-off}, giving explicit quantitative relations between communication-layer resource allocation in LAWNs and the achievable quality of closed-loop control. We first consider the fully observed case, where the encoder has direct access to the state \(\mathbf{x}_{k,t}\) without noise impact . Let \(\mathbf{S}_k\) be the positive-definite solution of the algebraic Riccati equation
\begin{equation}
\begin{aligned}
        \mathbf{S}_k
   & = \mathbf{Q}_k
      + \mathbf{A}_k^{\mathsf T}(\mathbf{S}_k - \mathbf{M}_k)\mathbf{A}_k,
    \\
    \mathbf{M}_k
    &\triangleq
   \mathbf{S}_k\mathbf{B}_k\big(\mathbf{R}_k + \mathbf{B}_k^{\mathsf T}\mathbf{S}_k\mathbf{B}_k\big)^{-1}\mathbf{B}_k^{\mathsf T}\mathbf{S}_k.
    \label{eq:ARE_full_k}
\end{aligned}
\end{equation}
In the absence of communication constraints, the minimal infinite-horizon LQR cost for UAV $k$ is
\begin{equation}
    \bar{J}_{k,\min} = \operatorname{tr}\big(\boldsymbol{\Sigma}_{\mathrm{v},k} \mathbf{S}_k\big),
\end{equation}
which corresponds to the idealized regime of perfect (infinite-rate) state feedback. For any admissible target average cost $\bar{J}_k > \bar{J}_{k,\min}$, it has been shown that every causal coding–control scheme must satisfy the fundamental lower bound \cite{kostina2019rate}
  \begin{equation}
    R_k
    \;\ge\;
    \log\big|\det \mathbf{A}_k\big|
    + \frac{n_1}{2}\log\!\left(
        1
        +  \frac{|\det(\boldsymbol{\Sigma}_{\mathrm{v},k} \mathbf{M}_k)|^{1/n_1}}
               {(\bar{J}_k - \bar{J}_{k,\min})/n_1}
    \right).
    \label{eq:rate_cost_full_k}
\end{equation}
The first term $\log|\det \mathbf{A}_k|$ characterizes the fundamental rate needed to stabilize UAV $k$ in networked control. When the target control cost is loose, i.e., $\bar{J}_k \gg \bar{J}_{k,\min}$, the second logarithmic term in \eqref{eq:rate_cost_full_k} becomes negligible and the required rate approaches $\log|\det \mathbf{A}_k|$. In contrast, as the target cost is tightened and approaches its minimum value $\bar{J}_{k,\min}$, the denominator $(\bar{J}_k - \bar{J}_{k,\min})/n_1$ tends to zero, implying that the required rate becomes unbounded. Hence, driving the LQR cost arbitrarily close to $\bar{J}_{k,\min}$ demands arbitrarily large communication resources. 

When the process noise is non-Gaussian, the rate–control cost function can be written in a more general form as
\begin{equation}
    R_k
    \;\ge\;
    \log|\det \mathbf{A}_k|
    + \frac{n_1}{2}\log\!\left(
        1
        +  \frac{\mathcal{E}(\mathbf{v}_{k})|\det\mathbf{M}_k|^{1/n_1}}
               {(\bar{J}_k - \bar{J}_{k,\min})/n_1}
    \right),
    \label{eq:rate_cost_full_k1}
\end{equation}
where $\mathcal{E}(\mathbf{v}_k)$ denotes the entropy power of the $n_1$-dimensional process-noise vector $\mathbf{v}_k$, defined as
  $  \mathcal{E}(\mathbf{v}_k)
    \triangleq
    \frac{1}{2\pi e}
    \exp\!\left(\frac{2}{n_1} h(\mathbf{v}_k)\right),$
with $h(\mathbf{v}_k)$ the differential entropy of $\mathbf{v}_k$. The entropy power satisfies the classical bounds
   $ \mathcal{E}(\mathbf{v}_k)
    \;\le\;
    \big(\det \boldsymbol{\Sigma}_{\mathrm{v},k}\big)^{\frac{1}{n_1}}
    \;\le\;
    \frac{1}{n_1}\operatorname{tr}\!\big(\boldsymbol{\Sigma}_{\mathrm{v},k}\big),$
with equality in the first inequality if and only if $\mathbf{v}_k$ is Gaussian and in the second inequality if and only if $\mathbf{v}_k$ is white. When $\mathbf{v}_k$ is Gaussian with covariance $\boldsymbol{\Sigma}_{\mathrm{v},k}$, one has $\mathcal{E}(\mathbf{v}_k) = \big(\det \boldsymbol{\Sigma}_{\mathrm{v},k}\big)^{1/n_1}$ and the bound in \eqref{eq:rate_cost_full_k1} reduces to \eqref{eq:rate_cost_full_k}.

In practice, the BS only has access to a noisy and partial observation of the state of UAV $k$. In this setting, the optimal structure obeys a separation principle: a steady-state Kalman filter first produces a state estimate, which is then fed to an LQR controller. Let $\boldsymbol{\Sigma}_k$ denote the steady-state error covariance of the Kalman filter for UAV $k$. The minimum achievable infinite-horizon LQR cost in the absence of communication constraints is
\begin{equation}
    \bar{J}_{k,\min}
    = \operatorname{tr}(\boldsymbol{\Sigma}_{\mathrm{v},k} \mathbf{S}_k)
      + \operatorname{tr}\!\big(\boldsymbol{\Sigma}_k\,\mathbf{A}_k^{\mathsf T}\mathbf{M}_k\mathbf{A}_k\big),
\end{equation}
where $\mathbf{S}_k$ and $\mathbf{M}_k$ are still given by \eqref{eq:ARE_full_k}. The second term represents the additional cost induced by estimation errors. Define
\begin{equation}
    \mathbf{N}_k
    \triangleq
    \mathbf{A}_k\boldsymbol{\Sigma}_k\mathbf{A}_k^{\mathsf T}
      - \boldsymbol{\Sigma}_k
      + \boldsymbol{\Sigma}_{\mathrm{v},k},
    \label{eq:N_partial_k}
\end{equation}
which characterizes the effective process noise seen by the controller after filtering. The rate–cost trade-off in the partially observed case is then given by
\begin{equation}
    R_k
    \;\ge\;
    \log\|\det\mathbf{A}_k\|
    + \frac{n_1}{2}\log\!\left(
        1
        +  \frac{|\det(\mathbf{N}_k\mathbf{M}_k)|^{1/n_1}}
               {(\bar{J}_k - \bar{J}_{k,\min})/n_1}
    \right).
    \label{eq:rate_cost_partial_k}
\end{equation}
Compared with \eqref{eq:rate_cost_full_k}, the matrix $\mathbf{N}_k$ incorporates the extra uncertainty created by observation noise and filtering. For the same target cost $\bar{J}_k$, the required rate in the partially observed case is therefore strictly larger, which reflects the additional information that must be communicated in order to reconstruct the state of UAV $k$ with sufficient accuracy. Given the plant matrices and noise statistics $\{\mathbf{A}_k,\mathbf{B}_k,\boldsymbol{\Sigma}_{\mathrm{v},k},\boldsymbol{\Sigma}_{\mathrm{w},k}\}$, the bounds in \eqref{eq:rate_cost_full_k} and \eqref{eq:rate_cost_partial_k} map a desired control cost level $\bar{J}_k$ to a minimum rate requirement $R_k$ for LAWN system design.

\subsubsection{Outage-Constrained Predictive Control}

The rate–cost bounds in the previous subsection describe the {average} information required to achieve a given LQR performance. In practice, LAWN-based control loops operate over time-varying wireless channels, where even if the long-term rate budget is sufficient according to \eqref{eq:rate_cost_full_k}–\eqref{eq:rate_cost_partial_k}, individual control packets may still be lost due to instantaneous channel fades or interference. When the control command at time $t$ is not successfully decoded, the UAV cannot apply the newly optimized input and typically resorts to holding the previous command. This motivates an outage-constrained predictive control framework, in which the communication unreliability is explicitly accounted for in both the controller design and the resource allocation strategy.

Model predictive control (MPC) provides a natural framework in this regard. At each instant $t$, the controller observes the current state $\mathbf{x}_{k,t}$ and solves a finite-horizon optimal control problem over a prediction horizon $N_p$ based on the transition model. This yields a sequence of candidate future inputs $\{\mathbf{u}_{k,t|t},\mathbf{u}_{k,t+1|t},\ldots,\mathbf{u}_{k,t+N_p-1|t}\},$
where the first element $\mathbf{u}_{k,t|t}$ is applied at time $t$, while the remaining elements serve as a predictive strategy for subsequent steps. At the next slot, the horizon is shifted forward and the optimization is repeated using the updated state, leading to the classical receding-horizon implementation \cite{jin2025mpc}.

When the control commands are conveyed over a fading channel, the actuator does not necessarily receive $\mathbf{u}_{k,t|t}$ at every time instant. A common actuation policy in this context is the \emph{hold-input} rule: if a packet is lost, the UAV continues to apply the most recently received command. Let $\delta_{k,t}\in\{0,1\}$ denote the reception indicator at time $t$, where $\delta_{k,t}=1$ if the packet is successfully decoded and $\delta_{k,t}=0$ otherwise. The actual input applied to UAV~$k$ can then be written as
\begin{equation}
    \tilde{\mathbf{u}}_{k,t}
    =
    \delta_{k,t}\,\mathbf{u}_{k,t|t}
    +
    (1-\delta_{k,t})\,\tilde{\mathbf{u}}_{k,t-1},
    \label{eq:u_hold_rule}
\end{equation}
which shows that packet outages induce a random input-holding mechanism. Substituting \eqref{eq:u_hold_rule} into the plant dynamics yields the stochastic closed-loop model
\begin{equation}
    \mathbf{x}_{k,t+1}
    =
    \mathbf{A}_k\mathbf{x}_{k,t}
    +
    \mathbf{B}_k
    \big(
        \delta_{k,t}\mathbf{u}_{k,t|t}
        +
        (1-\delta_{k,t})\tilde{\mathbf{u}}_{k,t-1}
    \big)
    +
    \mathbf{v}_{k,t},
    \label{eq:state_outage}
\end{equation}
where the packet-drop process $\{\delta_{k,t}\}$ appears as a multiplicative disturbance governed by the wireless link. The impact of the communication is captured through the outage probability. Let $p_{k,t}^{\mathrm{out}}\triangleq \Pr\{\delta_{k,t}=0\}$ denote the probability that the control packet for UAV~$k$ is lost at time $t$. It is evident that $p_{k,t}^{\mathrm{out}}$ is determined by the instantaneous SINR and the selected coding rate, i.e., $p_{k,t}^{\mathrm{out}}=\Pr\big\{R_{k,t}<R_{k}^{\mathrm{ctl}}\big\},$
where $R_{k}^{\mathrm{ctl}}$ is the required communication rate for transmitting a control packet. Consequently, the dual-functional beamforming directly shapes the statistics of $\{\delta_{k,t}\}$ and thus the effective dynamics {in \eqref{eq:state_outage}}. Within this setting, a standard quadratic finite-horizon cost takes the form
\begin{equation}
    J_{k,t}=
        \sum_{j=0}^{N_p}
        \big(
            \|\mathbf{x}_{k,t+j|t}-\mathbf{x}^{\mathrm{ref}}_{k,t+j}\|_{\mathbf{Q}_k}^2
            +
            \|\mathbf{u}_{k,t+j|t}\|_{\mathbf{R}_k}^2
        \big).
    \label{eq:mpc_cost_expectation}
\end{equation}
The control design at time $t$ then involves solving the quadratic minimization problem described above. The MPC formulation implicitly assumes that the ISAC transmitter has perfect knowledge of the downlink channels when designing the beamformers and coding rates that determine $p_{k,t}^{\mathrm{out}}$. In practice, however, the BS or controller only has access to imperfect channel estimates. A common model is given by
\begin{equation}
    \mathbf{h}_{k,t}
    = \hat{\mathbf{h}}_{k,t} + \Delta\mathbf{h}_{k,t},
    \label{eq:csit_error_model}
\end{equation}
where $\hat{\mathbf{h}}_{k,t}$ denotes the estimated channel from the BS to UAV $k$, and $\Delta\mathbf{h}_{k,t}$ models the channel-estimation error. In general, two canonical uncertainty models are of particular interest.
\begin{itemize}
    \item {Statistical CSI error model\cite{le2013downlink}:}
The error $\Delta\mathbf{h}_{k,t}$ is treated as a zero-mean CSCG random vector
\begin{equation}
    \Delta\mathbf{h}_{k,t}
    \sim
    \mathcal{CN}\!\left(
        \mathbf{0},
        \sigma^{2}_{h_k}\mathbf{I}_{N_t}
    \right),
    \label{eq:statistical_csit_error}
\end{equation}
where $\sigma^{2}_{h_k}$ is the variance of the estimation error and $N_t$ is the number of transmit antennas.  

\item{Bounded CSI error model\cite{zheng2009robust}:}
Alternatively, the error vector is assumed to belong to a deterministic compact set  
\begin{equation}
    \mathcal{H}_{k,t}
    \triangleq
    \big\{
        \mathbf{h}_{k,t}
        = \hat{\mathbf{h}}_{k,t} + \Delta\mathbf{h}_{k,t}
        \;:\;
        \|\Delta\mathbf{h}_{k,t}\|_2 \le \varepsilon_{k,t}
    \big\}, \label{eq:bounded_csit_error}
\end{equation}%
where $\varepsilon_{k,t}$ represents the radius of the uncertainty region.
\end{itemize} 
Under imperfect CSI, the control packet outage probability at time $t$ depends jointly on small–scale fading and channel uncertainty\cite{wu2025sagin}. Then, the corresponding robust design aims to enforce a chance constraint of the form  
\begin{equation}
    \Pr_{\Delta\mathbf{h}_{k,t}}
    \big\{
        R_{k,t}(\hat{\mathbf{h}}_{k,t}+\Delta\mathbf{h}_{k,t})
        < R_{k,t}^{\mathrm{ctl}}
    \big\}
    \le
    \kappa^{\mathrm{ctl}}_{k,t},
\end{equation}
ensuring that the control-oriented rate constraint is violated with probability at most $\kappa^{\mathrm{ctl}}_{k,t}$ under the given error statistics. 

To handle chance constraints induced by \eqref{eq:statistical_csit_error}, a complex Bernstein-type inequality is particularly useful.

\medskip
\noindent\textbf{Lemma 1 (Bernstein-type inequality \cite{Bernstein})}
Let $\mathbf{t}\sim\mathcal{CN}(\mathbf{0},\mathbf{I}_M)$,
$\mathbf{Q}\in\mathbb{H}^{M\times M}$ (Hermitian), and
$\mathbf{u}\in\mathbb{C}^{M\times 1}$. Consider the quadratic random
variable
  $  f(\mathbf{t}) = \mathbf{t}^{\mathsf H}\mathbf{Q}\mathbf{t}
                    + 2\Re\{\mathbf{t}^{\mathsf H}\mathbf{u}\}.$
For any $\kappa>0$, define $   \Upsilon(\kappa)
    \triangleq
    \operatorname{tr}(\mathbf{Q})
    - \sqrt{2\kappa}\,\sigma_f
    - \kappa\,\tau_f,$
where $  \sigma_f^2
    \triangleq \|\mathbf{Q}\|_{\mathrm{F}}^{2} + 2\|\mathbf{u}\|_2^{2}$
 and 
  $  \tau_f
    \triangleq \max\big\{\lambda_{\max}(\mathbf{Q}),\,0\big\},$ where $\lambda_{\max}(\cdot)$ denotes the maximum eigenvalue of a matrix. 
Then, the following probabilistic bound holds:
\begin{equation}
    \Pr\big\{ f(\mathbf{t}) \ge \Upsilon(\kappa) \big\}
    \;\ge\; 1 - e^{-\kappa}.
    \label{eq:bernstein_inequality}
\end{equation}

In robust beamforming with statistical CSI errors, the  rate for UAV $k$ can be rewritten as a quadratic inequality in
$\Delta\mathbf{h}_{k,t}$, which has the form of $f(\mathbf{t})$ with
$\mathbf{t} = \Delta\mathbf{h}_{k,t}/\sigma_{h,k}$, and suitable
$\mathbf{Q},\mathbf{u}$ determined by the beamformers. By choosing
$\kappa = -\ln(1-\rho_{k,t})$ for a required reliability level
$\rho_{k,t}=1-\kappa^{\mathrm{ctl}}_{k,t}$, the chance constraint is given by
\begin{align}
    \Pr\!\big\{
        R_{k,t}(\hat{\mathbf{h}}_{k,t}+\Delta\mathbf{h}_{k,t})
        \ge R_{k,t}^{\mathrm{ctl}}
    \big\}
    \;\ge\; \rho_{k,t},
    \end{align}
which is conservatively enforced by requiring that the deterministic inequality $f(\mathbf{t})\ge \Upsilon(\kappa)$ holds, becoming a set of linear and second-order cone constraints in the beamforming variables. In this way, the probabilistic outage constraint under statistical CSI errors is converted into a convex deterministic constraint that can be embedded into the outage-aware predictive control and ISAC beamforming design.

For norm-bounded CSI errors, one may optimize the system performance with worst-case QoS constraints. After standard manipulations, \eqref{eq:bounded_csit_error} can be expressed as a quadratic inequality in $\Delta\mathbf{h}_{k,t}$ that must hold for all $\Delta\mathbf{h}_{k,t}$ satisfying $\|\Delta\mathbf{h}_{k,t}\|_2^{2}\le \varepsilon_{k,t}^{2}$. The S-procedure gives an equivalent LMI condition for such implications.

\medskip
\noindent\textbf{Lemma 2 (S-procedure \cite{Sprocedure})}
Let
\begin{align}  
\phi_i(\mathbf{x})
   & = \mathbf{x}^{\mathsf H}\mathbf{A}_i\mathbf{x}
      + 2\Re\{\mathbf{b}_i^{\mathsf H}\mathbf{x}\}
      + c_i,
    \qquad i\in\{1,2\},
\end{align}
\textrm{with} $\mathbf{A}_i\in\mathbb{H}^{M\times M}$,~$\mathbf{b}_i\in\mathbb{C}^{M\times 1},~\textrm{and}~c_i\in\mathbb{R}$. Assuming that there exists some $\bar{\mathbf{x}}$ such that $\phi_1(\bar{\mathbf{x}})<0$, the implication
\begin{equation}
    \phi_1(\mathbf{x}) \le 0
    \ \Rightarrow\
    \phi_2(\mathbf{x}) \le 0
\end{equation}
holds if and only if there exists a scalar $\lambda\ge 0$ such that
\begin{equation}
    \begin{bmatrix}
        \mathbf{A}_2 & \mathbf{b}_2 \\
        \mathbf{b}_2^{\mathsf H} & c_2
    \end{bmatrix}
    +
    \lambda
    \begin{bmatrix}
        \mathbf{A}_1 & \mathbf{b}_1 \\
        \mathbf{b}_1^{\mathsf H} & c_1
    \end{bmatrix}
    \preceq \mathbf{0}.
    \label{eq:S_procedure_LMI}
\end{equation}

To apply Lemma~2, we first express the uncertainty set $|\Delta\mathbf{h}{k,t}|2^{2}\le \varepsilon{k,t}^{2}$ in the form $\phi_1(\Delta\mathbf{h}{k,t})\le 0$, with
$\mathbf{A}_1=\mathbf{I},\ \mathbf{b}_1=\mathbf{0},\ c_1=-\varepsilon_{k,t}^{2}$. Similarly, the worst-case rate constraint in \eqref{eq:bounded_csit_error} is written as $\phi_2(\Delta\mathbf{h}_{k,t})\le 0$ with appropriately chosen $\mathbf{A}_2,\mathbf{b}2,c_2$ determined by the beamformers and the rate threshold $R{k,t}^{\mathrm{ctl}}$. The S-procedure then replaces the resulting semi-infinite constraint \eqref{eq:bounded_csit_error} by the single LMI \eqref{eq:S_procedure_LMI}, augmented by a nonnegative scalar multiplier $\lambda$. This LMI can be handled efficiently via semidefinite programming. Together, Lemma~1 and Lemma~2 provide two complementary tools for enforcing control-driven reliability constraints under statistical and norm-bounded CSI errors, respectively, while preserving the tractability of the overall co-design problem.
 
\subsubsection{State Estimation with Imperfect Observations}

In LAWN-based remote control, the controller does not directly observe the true plant state $\mathbf{x}_{k,t}$ but instead relies on noisy measurements acquired at a sensing node. 
Relying on KF and its variants discussed in Sec. III-C, the estimated state $\tilde{\mathbf{x}}_{k,t}$ can be obtained at the sensing plane. Through a communication link, the controller can acquire this estimate. As a result, the controller-side state is additionally degraded by communication-induced distortion. This entire procedure follows the Markov chain
$\mathbf{x}_{k,t} \rightarrow \tilde{\mathbf{x}}_{k,t} \rightarrow \hat{\mathbf{x}}_{k,t}.$ 
The state estimation process produces an estimated state $\tilde{\mathbf{x}}_{k,t}$ with sensing distortion $D_{\mathrm{s}} = \mathbb{E}\!\left[\|\mathbf{x}_{k,t}-\tilde{\mathbf{x}}_{k,t}\|^2\right]$, while the communication process, which transmits $\tilde{\mathbf{x}}_{k,t}$ to the controller, produces an additional recovery distortion $D_{\mathrm{c}} = \mathbb{E}\!\left[\|\tilde{\mathbf{x}}_{k,t}-\hat{\mathbf{x}}_{k,t}\|^2\right]$.
The total distortion can be expressed as
\begin{equation}
D = \mathbb{E}\left[\|\mathbf{x}_{k,t}-\hat{\mathbf{x}}_{k,t}\|^2\right],
\end{equation}
which explicitly shows how sensing errors and transmission errors jointly determine the reliability of closed-loop LAWN control. To characterize $D_{\mathrm{c}}$, one can adopt the rate–distortion (RD) formulation. Let the estimated state $\tilde{\mathbf{x}}_{k,t}$ follow a Gaussian distribution with covariance $\mathbf{R}_{\tilde{x}}$. For a target recovery distortion $D_{\mathrm{c}}$, RD theory states that the minimum coding rate required for reliable transmission is \cite{dong2025communication}
\begin{equation}
R(D_{\mathrm{c}})=
\sum_{i} \log\!\left(\frac{\lambda_i(\mathbf{R}_{\tilde{x}})}{D_i}\right),
\end{equation}
where $\lambda_i(\mathbf{R}_{\tilde{x}})$ are the eigenvalues of the estimate covariance and $D_i$ are per-mode distortions governed by the reverse water-filling rule
\begin{equation}
D_i = \lambda_i(\mathbf{R}_{\tilde{x}}) - \big(\lambda_i(\mathbf{R}_{\tilde{x}})-\xi\big)^{+}.
\end{equation}
The parameter $\xi$ is determined such that $\sum_i D_i = D_{\mathrm{c}}$. Communication distortion is therefore jointly determined by the uncertainty of the estimated state and the communication channel capacity $C$. According to the source-channel separation theorem, $R(D_{\mathrm{c}}) \le C$ is necessary and sufficient for achieving distortion $D_{\mathrm{c}}$, which shows directly how limited channel capacity increases the error in the recovered state $\hat{\mathbf{x}}_{k,t}$. The RD formulation provides a precise framework to interpret the coupling among sensing, communication, and control in LAWNs. For a fixed sensing strategy, the sensing distortion $D_{\mathrm{s}}$ describes the fidelity of the intermediate estimate $\tilde{\mathbf{x}}_{\mathrm{e}}$ and, consequently, the amount of information that must be conveyed to the controller. RD theory then implies that representing a more accurate estimate at the controller generally requires a higher communication rate in order to keep the additional communication-induced distortion $D_{\mathrm{c}}$ within acceptable limits. Conversely, when the available rate is constrained, stronger compression becomes unavoidable, which increases $D_{\mathrm{c}}$ and degrades the quality of the recovered parameter $\hat{\mathbf{x}}$, thereby tightening the margin for closed-loop stability and control performance. This viewpoint underscores that sensing waveform design, source–channel coding, and control policy design cannot be optimized in isolation. Sensing accuracy must be chosen with the downstream rate budget and the communication strategy must reflect the control task’s tolerance to estimation error.

\subsubsection{Energy-Aware Control Design}
\begin{figure*}[h]
    \begin{equation}
    \bar{E}
    =
    \int_{0}^{T}
    \left(
      c_1 \|\mathbf{v}(t)\|^{3}
      +
      c_2 \frac{1}{\|\mathbf{v}(t)\|}
      \Bigg(
        1
        +
        \frac{\|\mathbf{a}(t)\|^{2}
        -
        \dfrac{\big(\mathbf{a}(t)^{\mathsf T}\mathbf{v}(t)\big)^{2}}
        {\|\mathbf{v}(t)\|^{2}}}{g^{2}}
      \Bigg)
    \right)\! dt
    +
    \frac{m}{2}
    \big(\|\mathbf{v}(T)\|^{2}-\|\mathbf{v}(0)\|^{2}\big).
    \label{eq:UAV_energy_model}
\end{equation}
\hrule
\end{figure*}
Building on the previous subsection, where imperfect observations and communication distortions were shown to affect the controller-side state, we now shift attention to the energy constraints that fundamentally shape LAWN operations. The power consumption of a UAV platform can be conceptually divided into two components. The first corresponds to the communication–sensing–control functionality, which includes RF transmission, reception, baseband processing, and sensing operations. These tasks typically consume only a few watts. The second component is the propulsion subsystem, which maintains flight and enables maneuvering in low-altitude airspace. Propulsion power often reaches hundreds of watts and therefore dominates the total energy budget. This high power disparity implies that energy-aware design in LAWNs must be driven primarily by the cost of UAV motion rather than by marginal variations in transmit or sensing power.

To understand how control policies and information-gathering strategies translate into energy cost, we consider a fixed-wing UAV flying at a constant altitude with horizontal trajectory \(\mathbf{q}(t)\), instantaneous velocity \(\mathbf{v}(t) = \dot{\mathbf{q}}(t)\), and acceleration \(\mathbf{a}(t) = \ddot{\mathbf{q}}(t)\). The propulsion power required to sustain aerodynamic lift and counteract drag can be expressed as a function of speed and maneuvering intensity\cite{zeng2017energy}. Over a time horizon \(T\), the total propulsion energy is given by \eqref{eq:UAV_energy_model}. Here, \(c_1\) and \(c_2\) are aerodynamic parameters of the airframe, \(m\) is the UAV mass, and \(g\) is the gravitational constant. The term \(c_1\|\mathbf{v}(t)\|^3\) reflects parasitic drag, which grows rapidly with speed, while the second term accounts for lift-induced drag and highlights the energetic cost of changes in heading. The kinetic-energy term captures the effect of net acceleration across the trajectory. This formulation reveals that both high-speed flight and aggressive turning significantly increase energy consumption, underscoring the need to integrate propulsion considerations directly into LAWN control strategies.

To model this effect, consider a fixed-wing UAV flying at constant altitude with speed \(V_0\) along the negative \(y\)-axis. The airflow over its wings generates a pair of semi-infinite trailing vortices with circulation \(\zeta\) and lateral separation \(\alpha = \beta\pi/4\), where \(\beta\) is the wingspan. Under the horseshoe-vortex approximation, the induced velocity at a downstream point \((x,y)\) can be expressed using a Biot–Savart-type integral. For tractable analysis in LAWNs, this induced field is commonly approximated by a NASA–Burnham–Hallock model with Gaussian decay along the streamwise direction, which yields
\begin{equation}
    v_{\mathrm{in}}(x,y)
    =
    \frac{\zeta}{2\pi}
    \frac{x}{r_c^2 + x^2}
    \left(
        1
        +
        \frac{y}{\sqrt{(\beta/2)^2 + y^2}}
    \right)
    e^{\left(
        -\frac{(y-\mu)^2}{2\sigma_0^2}
    \right)},
    \label{eq:vin_single_uav}
\end{equation}
where \(r_c\) is a core-radius parameter and \(\mu,\sigma_0\) control the longitudinal decay of the vortex. The vertical upwash experienced by a trailing UAV is obtained by averaging the induced velocity over its wingspan. Denoting by \(u_0(x,y)\) the average upwash at lateral offset \(x\) and longitudinal spacing \(y\), we have
\begin{equation}
    u_0(x,y)
    =
    \frac{1}{\beta}
    \int_{x-\beta/2}^{x+\beta/2}
        v_{\mathrm{in}}(\eta,y)\,\mathrm{d}\eta,
    \label{eq:avg_upwash}
\end{equation}
which admits a closed-form expression under \eqref{eq:vin_single_uav} and exhibits two dominant upwash lobes near the wingtips and a downwash region under the fuselage. Positive values of \(u_0(x,y)\) correspond to beneficial upwash that reduces the lift the follower must generate, whereas negative values correspond to downwash that increases induced drag.

The drag and power reductions for a follower UAV can then be quantified in terms of \(u_0(x,y)\). Let \(F_l\) denote the baseline lift required for level flight. The reduction in induced drag at each wing is
\begin{equation}
    \Delta F_d(x,y)
    =
    \frac{F_l\,u_0(x,y)}{V_0},
    \label{eq:drag_reduction}
\end{equation}
and the corresponding reduction in propulsion power is
\begin{equation}
    \Delta P(x,y)
    =
    F_l\,u_0(x,y).
    \label{eq:power_reduction}
\end{equation}
The expressions \eqref{eq:drag_reduction}–\eqref{eq:power_reduction} show that formation-induced energy savings are maximized when each follower maintains a relative position \((x,y)\) that maximizes the local upwash function \(u_0(x,y)\), typically at a lateral offset of roughly one wingspan and a longitudinal spacing of roughly one to two wingspans behind its reference UAV.

In a multi-UAV LAWN, each vehicle is influenced by the vortices generated by all others. Denote the horizontal position of the \(k\)-th UAV in a given formation by \(\mathbf{z}_k = [x_k, y_k]^{\mathsf T}\). The total upwash observed by UAV \(k\) from the remaining UAVs is then
\begin{equation}
    u_{\mathrm{tot}}(\mathbf{z}_k)
    =
    \sum_{k' \neq k}
    u_0\big(x_k - x_{k'},\, y_k - y_{k'}\big).
    \label{eq:total_upwash}
\end{equation}
These expressions show that, once the single-UAV propulsion power model has been established, the energy-aware design of a large-scale LAWN naturally extends to a cooperative optimization problem. The control objective is no longer restricted to tracking waypoints or maintaining connectivity. Instead, it also seeks UAV configurations \(\{\mathbf{z}_k\}\) that maximize \(u_{\mathrm{tot}}(\mathbf{z}_k)\) for all followers while respecting collision-avoidance and mission constraints, thereby converting aerodynamic interactions into tangible propulsion-power savings for the entire network.

So far, the energy-aware design of LAWNs has focused on reducing propulsion expenditure by shaping individual trajectories and exploiting aerodynamic formations. Even with these gains, however, the network lifetime is ultimately capped by the finite battery capacity carried on each platform. A natural complement to energy-efficient motion is therefore to replenish UAV energy wirelessly, reusing the existing ISAC downlink not only for communication and sensing but also for power transfer \cite{wei2021resource}. Based on the ISAC transmit model in \eqref{eq:isac-tx}–\eqref{eq:tx-cov}, the same downlink waveform can also be exploited for wireless power transfer, so that the BS simultaneously delivers control information, sensing probes, and RF energy to the UAV formations. Consider the leader of formation \(k\) at time $t$ with downlink channel \(\mathbf{h}_{k,t}\in\mathbb{C}^{1\times N_t}\). The received baseband signal is \(y_{k,t}=\mathbf{h}_{k,t}\mathbf{s}_n+z_{k,t}\), and, neglecting noise for energy-harvesting purposes, the corresponding average received RF power is
\begin{equation}
    P^{\mathrm{RF}}_{k,t}
    =
    \mathbb{E}\!\left[|y_{k,t}|^{2}\right]
    =
    \mathbf{h}_{k,t}\mathbf{C}_t\mathbf{h}_{k,t}^\mathsf{H},
    \label{eq:Pr-matrix-final}
\end{equation}
which shows that the transmit covariance \(\mathbf{C}_t\) not only determines the communication and sensing performance, but also controls the spatial distribution of RF energy in the LAWN.

Equipping each leader UAV with a power-splitting SWIPT receiver, a fraction \(\alpha_{k,t}\in[0,1]\) of the received power is routed to the information decoder for control, while the remaining fraction \(1-\alpha_{k,t}\) is delivered to the rectifier. Under the commonly used linear EH model, the harvested DC power is
\begin{equation}
    P^{\mathrm{lin}}_{e,k,t}
    =
    \eta_k\big(1-\alpha_{k,t}\big)\,
    P^{\mathrm{RF}}_{k,t}
    =
    \eta_k\big(1-\alpha_{k,t}\big)\,
    \mathbf{h}_{k,t}\mathbf{C}_t\mathbf{h}_{k,t}^\mathsf{H},
    \label{eq:EH-linear-final}
\end{equation}
where \(0<\eta_k\le 1\) denotes the RF–to–DC conversion efficiency. This model highlights the basic tradeoff between information decoding and energy harvesting, but it ignores the sensitivity and saturation of practical rectennas.

To capture these hardware effects, we can instead adopt a nonlinear logistic EH model. Let
$  P^{\mathrm{EH}}_{k,t}= \big(1-\alpha_{k,t}\big)\,
\mathbf{h}_{k,t}\mathbf{C}_t\mathbf{h}_{k,t}^\mathsf{H}$
denote the RF power impinging on the rectifier. The harvested DC power is then modeled as
\begin{equation}
    P^{\mathrm{sat}}_{e,k,t}
    =
    \frac{\Psi_{k,t}-P^{\mathrm{Sat}}_{k}\Omega_k}{1-\Omega_k},
    \label{eq:EH-sat-final}
\end{equation}
where
\begin{align}
    \Psi_{k,t}
        &=
        \frac{P^{\mathrm{Sat}}_{k}}
        {1+\exp\!\big(-a_k\big(P^{\mathrm{EH}}_{k,t}-b_k\big)\big)},\\
    \Omega_k
        &=
        \frac{1}{1+\exp(a_k b_k)}.
\end{align}
Here, \(P^{\mathrm{Sat}}_{k}\) is the saturation harvested power, \(a_k>0\) controls the slope of the charging curve, and \(b_k\) is the effective turn-on threshold. Compared with \eqref{eq:EH-linear-final}, the nonlinear model \eqref{eq:EH-sat-final} reproduces the low-input sensitivity, quasi-linear region, and high-power saturation observed in real EH circuits, while retaining a simple dependence on the covariance-induced RF power. Let \(\ell_{k,t}\) denote the instantaneous control-stage cost of formation \(k\), and let
$    P^{\mathrm{EH}}_{e,k,t}
    \in
    \left\{
        P^{\mathrm{lin}}_{e,k,t},
        \;P^{\mathrm{sat}}_{e,k,t}
    \right\}$
denote the harvested power under either EH model. Then, one possible energy-aware SWIPT design that explicitly balances control performance and wireless replenishment can be through maximizing $ {\sum_{t}\sum_{k} P^{\mathrm{EH}}_{e,k,t}/ \ell_{k,t}},$
while subject to the BS transmit power, the power-splitting constraints, and other required ISAC requirements. The formulation maximizes the ratio between harvested power and control cost and makes explicit how the ISAC covariance design and the SWIPT splitting factors jointly determine both the sustainability of LAWN energy supply and the quality of closed-loop control.

\section{AI for LAWNs}

Signal processing provides the physical-layer and estimation building blocks, but a large-scale LAWN must also decide, coordinate, and adapt across time scales and across many agents. Building upon the signal processing foundations established in the previous section, AI emerges as a transformative paradigm that enables adaptive, autonomous, and intelligent decision-making in LAWNs. Specifically, the integration of AI technologies addresses the complex challenges inherent in dynamic 3D airspace operations, including real-time adaptation to environmental changes, multi-objective optimization under resource constraints, and coordinated behavior among heterogeneous aerial platforms. Moreover, the capacity of AI for learning from experience and predicting future states makes it particularly well-suited for the highly dynamic and uncertain operational environments characteristic of low-altitude airspace. Consequently, this section explores the fundamental AI concepts relevant to LAWNs, examines how AI enhances core network functionalities, and investigates emerging AI paradigms that promise to revolutionize autonomous aerial operations.

\subsection{AI Fundamentals for LAWNs}
\label{sec:ai-fundamentals}

\par The foundational AI technologies applicable to LAWNs can be broadly categorized into discriminative and generative approaches, each serving distinct but complementary roles in enabling intelligent aerial network operations. Specifically, these fundamental AI paradigms provide the theoretical and practical basis for addressing challenges ranging from perception and decision-making to prediction and content generation. As such, understanding these core AI concepts is essential for developing effective solutions that leverage machine learning capabilities while respecting the computational, energy, and real-time constraints inherent in aerial platforms.

\subsubsection{Discriminative AI}

\par Discriminative AI models focus on learning decision boundaries and classification functions that enable UAVs to make informed decisions based on observed data, thus forming the backbone of perception and control systems in LAWNs. Specifically, these models excel at pattern recognition, classification, and regression tasks that are fundamental to autonomous navigation, obstacle detection, and network state assessment~\cite{Tang2023}. Moreover, discriminative approaches are particularly valuable in LAWNs due to their ability to process multi-modal sensor data and extract actionable insights in real-time~\cite{Xie2025}.

\par For instance, CNNs have evolved significantly from basic architectures to sophisticated attention-enhanced models that enable robust computer vision capabilities for aerial platforms. Initially, CNN architectures such as ResNet, MobileNet, and EfficientNet were adapted for onboard processing, providing foundational object detection and semantic segmentation with optimized computational efficiency for resource-constrained UAV platforms~\cite{Ma2025}. Subsequently, the integration of attention mechanisms and multi-scale feature extraction improved the ability to process visual data from cameras and LiDAR sensors for obstacle identification and terrain classification. Additionally, recent developments like SpectralMamba architectures utilize state-space modeling to achieve lightweight yet highly efficient feature extraction, obtaining superior classification accuracy while significantly reducing computational costs compared to traditional approaches~\cite{Yao2024}.

\par Likewise, RL algorithms have progressed from basic value-based methods to sophisticated distributional and meta-learning approaches that enable UAVs to learn optimal control policies in dynamic low-altitude environments. Early model-free RL approaches such as deep Q-networks (DQN), proximal policy optimization (PPO), and soft actor-critic (SAC) demonstrated initial success in trajectory planning, collision avoidance, and resource allocation tasks for individual UAVs~\cite{Ghomri2024}. Subsequently, MARL frameworks extended these capabilities to swarm scenarios, with algorithms like multi-agent deep deterministic policy gradient (MADDPG) enabling coordinated actions among multiple UAVs learning from collective experience~\cite{He2023}. \bluechange{Moreover, advanced distributional RL methods such as truncated quantile critics (TQC) have demonstrated the ability to handle uncertain returns in complex sequential decision problems~\cite{Xiao2023}. This characteristic is also relevant to resource allocation and control under time-varying LAWN conditions.}

\par
\begin{figure}[tb]
  \centering
  \includegraphics[width=0.4\textwidth]{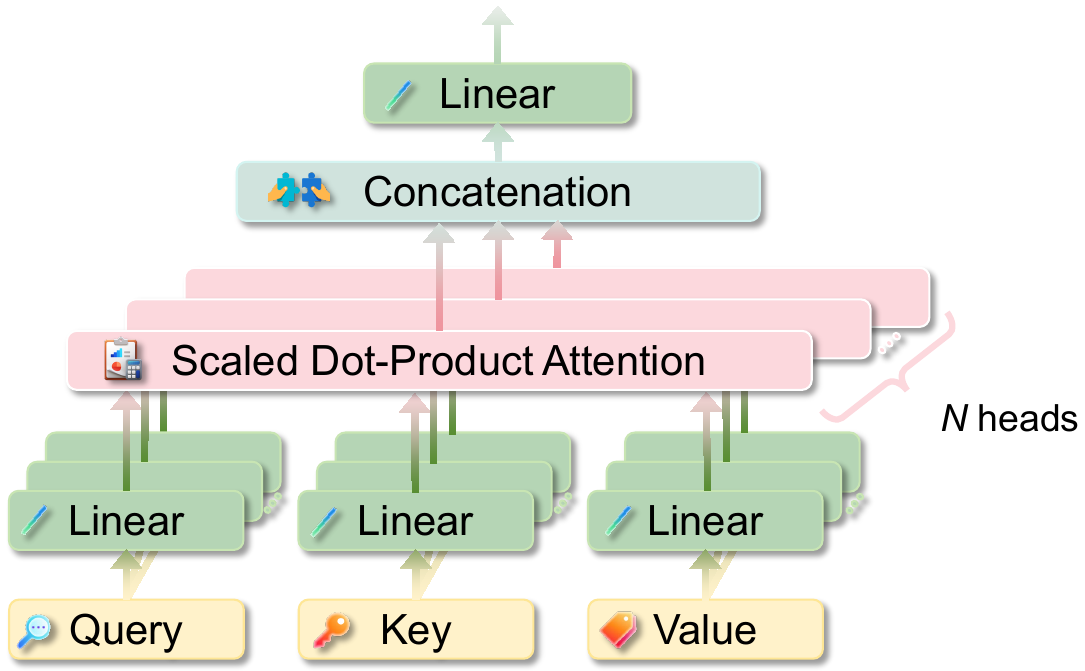}
  \caption{A diagram illustrating the multi-head attention mechanism.}
  \label{fig:multi-head}
\end{figure}

 Simultaneously, classical machine learning methods have evolved through integration with deep learning architectures to provide enhanced discriminative capabilities for RF signal analysis and network security applications in LAWNs. \bluechange{Initially, SVM-based methods provided interpretable decision boundaries for high-dimensional spectral data~\cite{Li2015}. Their relatively transparent decision process also makes them attractive for safety-critical signal classification tasks. Additionally, transformer-based intrusion detection has been explored in other cyber-physical networks~\cite{Zhao2025}. These studies provide a related design reference for LAWN security.} In particular, the attention mechanism in transformer-based LAWN security systems can be formulated as follows:
\begin{equation}
    \text{Attention}(Q, K, V) = \text{softmax}\left(\frac{QK^\mathsf{T}}{\sqrt{d_k}}\right)V,
\end{equation}
\noindent where $Q$, $K$, and $V$ are query, key, and value matrices derived from input features (e.g., RF signal properties or network traffic patterns), and $d_k$ is the dimension of the key vectors. This mechanism allows the model to focus on the most relevant parts of the input when making security-critical decisions. Likewise, as shown in Fig.~\ref{fig:multi-head}, multi-head attention further enhances this capability by computing multiple attention functions in parallel, \textit{i.e.,}
\begin{equation}
    \text{MultiHead}(X) = \text{Concat}(\text{head}_1, \text{head}_2, ..., \text{head}_h)W^O, 
\end{equation}
\noindent where each head $\text{head}_i = \text{Attention}(XW_i^Q, XW_i^K, XW_i^V)$ captures different relationship patterns in the input data $X$. Consequently, attention-enhanced ensemble methods and multi-scale fusion networks now enable robust detection and classification in the complex environments typical of LAWN deployments.

\subsubsection{Generative AI}

\par Generative AI models demonstrate compatibility with LAWN architectures by addressing the fundamental challenges of dynamic 3D airspace operations, including data scarcity, scenario diversity, and predictive uncertainty in LAWNs~\cite{Sun2025}. \bluechange{Specifically, the time-varying trajectories and resource conditions considered in predictive LAWN control create a demand for models that can anticipate future network states~\cite{Jin2025}. This demand also motivates generative modeling for scenario forecasting and adaptation.}

\par In particular, early generative architectures have evolved from basic variational autoencoders (VAEs) to sophisticated adversarial frameworks that are particularly well-suited for LAWN's multi-modal data requirements. Initially, VAEs provided foundational capabilities for learning compact latent representations of flight trajectories, channel conditions, and network topologies, enabling efficient compression and reconstruction of high-dimensional operational data in resource-constrained aerial environments~\cite{Xia2022}. Specifically, VAEs for LAWN applications optimize a dual objective function that balances reconstruction quality with distribution matching:
\begin{equation}
    \mathcal{L}(\theta,\phi;x) = \mathbb{E}_{q_\phi(z|x)}[\log p_\theta(x|z)] - D_{KL}(q_\phi(z|x)||p(z)), 
\end{equation}
\noindent where $q_\phi(z|x)$ is the encoder network with parameters $\phi$ mapping input data $x$ (such as flight trajectories or network states) to latent representations $z$, $p_\theta(x|z)$ is the decoder with parameters $\theta$ reconstructing the input, and $D_{KL}$ is the Kullback-Leibler divergence ensuring the learned latent distribution approximates a prior $p(z)$.

\par Subsequently, generative adversarial networks (GANs) revolutionized synthetic data generation by creating realistic datasets for training discriminative models when real-world LAWN data was limited or expensive to collect~\cite{Zhang2022}. In detail, the GAN framework operates through a minimax optimization between a generator $G$ and a discriminator $D$:
\begin{equation}
    \begin{aligned}
    \min_G \max_D V(D, G) &= \mathbb{E}_{x \sim p_{data}(x)}[\log D(x)] \\
    &+ \mathbb{E}_{z \sim p_z(z)}[\log(1 - D(G(z)))], 
    \end{aligned}
\end{equation}
\noindent where $p_{data}$ represents the real LAWN data distribution, $z$ is random noise, $G(z)$ generates synthetic samples, and $D(x)$ estimates the probability of data being real. Additionally, recent developments in attention-enhanced GANs and spectral normalization techniques have improved training stability and generation quality for complex 3D airspace scenarios~\cite{Chattopadhyay2022}.

\par Likewise, probabilistic generative models have advanced from classical Gaussian processes to sophisticated hierarchical frameworks that excel at uncertainty quantification in LAWNs. Initially, Gaussian processes (GPs) provided robust probabilistic modeling of channel gain maps, interference patterns, and environmental conditions with explicit uncertainty bounds, enabling risk-aware decision-making crucial for safe low-altitude operations~\cite{Englezou2024}. Moreover, deep GP architectures and neural process models have demonstrated superior performance in modeling complex spatial-temporal dependencies in multi-UAV networks. Consequently, modern Bayesian neural networks and uncertainty-aware deep learning frameworks now provide principled approaches to handling the inherent stochasticity of LAWN environments~\cite{Zhou2025}. These networks represent parameters as probability distributions, which are given by
\begin{equation}
    p(w|D) = \frac{p(D|w)p(w)}{p(D)}, 
\end{equation}
\noindent where $w$ represents network weights, $D$ is the observed data, $p(w)$ is the prior distribution over weights, and $p(D|w)$ is the likelihood. Following this, predictions incorporate uncertainty by integrating over the posterior weight distribution, \textit{i.e.},
\begin{equation}
    p(y|x,D) = \int p(y|x,w)p(w|D)dw, 
\end{equation}
\noindent where $y$ is the prediction for input $x$.

\par Simultaneously, sequence generation models have progressed from basic recurrent architectures to advanced diffusion models that are exceptionally well-matched to the temporal prediction requirements of LAWN. Initially, recurrent neural networks (RNNs) and their variants, including long short-term memory (LSTM) and gated recurrent unit (GRU) networks, enabled basic temporal sequence generation for forecasting network traffic patterns and channel quality evolution. Subsequently, attention-based transformer architectures demonstrated superior performance in long-horizon predictions of network behavior and environmental conditions, particularly valuable for LAWN mission planning~\cite{Zhu2023}. Furthermore, the emergence of diffusion models has revolutionized sequence generation by providing stable, high-quality synthesis of complex temporal patterns, including UAV trajectory sequences, interference evolution, and multi-modal sensor data streams~\cite{Zhang2025}. These models employ a forward process that gradually adds noise to data $x_0$ as follows:
\begin{equation}
    q(x_t|x_{t-1}) = \mathcal{N}(x_t; \sqrt{1-\beta_t}x_{t-1}, \beta_t\mathbf{I}), 
\end{equation}
\noindent where $x_0$ represents the original clean data (e.g., UAV trajectories), $x_t$ is the noised data at time step $t$, $\beta_t$ is the noise schedule parameter, and $\mathcal{N}$ denotes a normal distribution. The generative process then reverses this corruption through learned denoising, \textit{i.e.},
\begin{equation}
    p_\theta(x_{t-1}|x_t) = \mathcal{N}(x_{t-1}; \mu_\theta(x_t, t), \Sigma_\theta(x_t, t)), 
\end{equation}
\noindent where $\mu_\theta$ and $\Sigma_\theta$ are neural network predicted parameters controlling the reverse denoising step, and $\theta$ represents the learnable network weights. \bluechange{Additionally, diffusion-based restoration methods have reconstructed missing information in satellite imagery while preserving useful spatial structures under incomplete observations~\cite{Donguk2025}.}

%
\subsection{AI-Enhanced LAWN Functionalities}

\par Building upon the mentioned discriminative and generative AI foundations, the practical integration of these technologies significantly enhances the core functionalities of LAWNs, transforming reactive systems into proactive, adaptive, and intelligent networks capable of autonomous operation. Specifically, AI-enhanced functionalities span three critical domains of LAWN operations, which are individual UAV control and navigation, network-wide resource allocation and optimization, and proactive security and privacy management~\cite{Wu2025JSAC}. Furthermore, these AI-driven enhancements enable LAWNs to achieve performance levels and autonomous operation capabilities that would be impossible with traditional rule-based or static optimization approaches.

\subsubsection{AI for UAV Control and Navigation}
\label{sssec: AI for CN}

\par AI-driven control and navigation systems fundamentally transform LAWN operations by enabling autonomous decision-making, adaptive flight control, and intelligent path planning that are essential for safe and efficient low-altitude airspace management~\cite{Liu2024}. Specifically, these AI-enhanced systems address the unique challenges of LAWN environments, including dynamic obstacle avoidance in cluttered urban settings, formation control in dense UAV swarms, and mission execution under highly uncertain atmospheric conditions. Moreover, AI-enhanced navigation capabilities extend far beyond traditional autopilot systems by incorporating learning-based adaptation to environmental variability, platform-specific flight characteristics, and evolving mission requirements in 3D airspace~\cite{Krishnan2024}.

\par Initially, classical control approaches provided basic autopilot functionality, but the integration of deep RL frameworks has revolutionized autonomous flight control by enabling UAVs to learn optimal policies that simultaneously balance multiple competing objectives critical to LAWN operations~\cite{Wang2025}. Subsequently, actor-critic algorithms demonstrated superior performance in learning continuous control actions for attitude, altitude, and velocity management while dynamically adapting to changing wind conditions, varying payload configurations, and potential system failures~\cite{NasrAzadani2022}. {In an actor-critic architecture, the actor network parameterized by $\theta$ directly maps state $s$ to an action or an action distribution. For a stochastic actor, an action is sampled according to
\begin{equation}
    a \sim \pi_{\theta}(\cdot|s),
\end{equation}
\noindent while the critic network parameterized by $w$ evaluates the action-value and state-value functions and computes the advantage estimate as follows:
\begin{equation}
    A_w(s,a) = Q_w(s,a) - V_w(s).
\end{equation}
The actor is then optimized through the policy gradient, \textit{i.e.},
\begin{equation}
    \nabla_{\theta}J(\theta) = \mathbb{E}_{s,a}[\nabla_{\theta} \log \pi_{\theta}(a|s) A_w(s,a)],
\end{equation}
\noindent where $s$ represents the state and $a$ denotes the control action.} Furthermore, hierarchical RL approaches have proven particularly effective for LAWN applications by decomposing complex navigation tasks into multiple operational levels, with high-level strategic planners optimizing corridor assignments and low-level tactical controllers executing precise collision avoidance maneuvers~\cite{Sun2025a}. Additionally, recent developments in distributional RL and meta-learning have enhanced the robustness of LAWN control policies under the high uncertainty and non-stationary conditions characteristic of dynamic airspace environments~\cite{Eldeeb2025}.

\par Simultaneously, multi-agent coordination algorithms have advanced from basic consensus protocols to sophisticated distributed intelligence frameworks that enable seamless swarm behaviors and collaborative mission execution across heterogeneous LAWN platforms~\cite{Xie2025a}. Initially, consensus-based algorithms provided foundational capabilities for formation maintenance and collision avoidance while allowing individual UAVs to adapt to local environmental conditions and operational constraints~\cite{Zhang2018}. Subsequently, distributed optimization techniques enabled more sophisticated coordination capabilities, allowing UAV swarms to dynamically reconfigure their spatial topology and task allocation strategies based on evolving mission requirements and changing environmental conditions~\cite{Gao2022}. Furthermore, recent developments in MARL and graph neural networks have enhanced swarm intelligence capabilities by enabling emergent coordination behaviors and adaptive task specialization~\cite{Yan2024}. The mathematical foundation of multi-agent coordination can be represented through a formal tuple. A multi-agent system is defined as
\begin{equation}
    \langle N, S, \{A_i\}_{i=1}^N, T, \{R_i\}_{i=1}^N, \{\Omega_i\}_{i=1}^N, \gamma \rangle, 
\end{equation}
\noindent where $N$ represents the total number of UAVs. The global state space is denoted by $S$. Each UAV $i$ has an action space $A_i$ and observation space $\Omega_i$. The transition dynamics are captured by $T$. Individual reward functions are represented by $R_i$. The discount factor $\gamma$ weights future rewards.

\par As such, each UAV should optimize its policy $\pi_i(a_i|o_i)$ based on local observations. This policy determines the probability of selecting action $a_i$ given observation $o_i$. In general, the decentralized execution with a centralized training approach addresses this coordination challenge. This method uses the joint action-value function as follows:
\begin{equation}
    Q_{tot}(\boldsymbol{\tau}, \mathbf{a}) = f_{\theta}\left(Q_1(o_1, a_1), Q_2(o_2, a_2), \ldots, Q_N(o_N, a_N)\right). 
\end{equation}
The function $Q_{{tot}}$ represents the total team value,
while $\boldsymbol{\tau}$ and $\mathbf{a}$ denote the joint trajectory
history and joint action, respectively. The mixing network $f_{\theta}$
combines the individual Q-values to support coordinated decision-making.
Consequently, modern AI-driven coordination frameworks can support
complex LAWN applications, including distributed sensing, collaborative
cargo transport, and coordinated emergency response \cite{Betalo2025}.
\revise{Despite these advances, scalable MARL deployment in LAWNs remains
constrained by communication overhead and policy-induced
non-stationarity. If each UAV exchanges a $q$-dimensional state/action
message with all other UAVs, the signaling overhead per coordination
round scales as $\mathcal{O}(N^{2}q)$. Restricting communication to
local neighbors reduces this overhead to
$\mathcal{O}(N\bar{\Delta}q)$, where $\bar{\Delta}$ denotes the average
communication degree. Event-triggered communication, message
compression, and hierarchical coordination can further reduce the
signaling burden, while centralized training with decentralized
execution, parameter sharing among homogeneous agents, and off-policy
correction can improve training stability under concurrent policy
updates.}

\subsubsection{AI for Resource Allocation and Optimization}

\par AI-driven resource allocation and optimization techniques fundamentally address the complex multi-dimensional challenges inherent in LAWN operations, where the dynamic 3D nature and unprecedented scale of the optimization space render traditional static approaches inadequate~\cite{Yang2021}. Consequently, these intelligent AI methods enable real-time adaptive management of critical LAWN resources, including spectrum allocation across multiple altitude layers, power control for energy-constrained aerial platforms, trajectory planning in congested airspace, and computational task scheduling across heterogeneous UAV networks~\cite{Yan2025}. Moreover, AI optimization frameworks excel at handling the non-convex, multi-objective problems characteristic of LAWN environments while continuously learning from operational data to improve future resource allocation decisions~\cite{Sun2024}.

\par Spectrum management in LAWNs has evolved from basic static allocation schemes to sophisticated machine learning-driven frameworks that dynamically optimize frequency resources across the complex 3D airspace topology. Initially, simple interference avoidance algorithms provided basic spectrum sharing capabilities, but the integration of deep neural networks has revolutionized spectrum allocation by learning to predict availability patterns and interference dynamics based on UAV mobility, geographical terrain features, and temporal traffic variations. Subsequently, distributed learning approaches such as federated learning have enabled coordinated spectrum sensing and allocation across multiple LAWN operators without requiring centralized coordination or compromising sensitive operational data~\cite{Li2024a}. Specifically, in the federated learning framework, each LAWN operator $i$ optimizes its local model, which is given by
\begin{equation}
    \min_{w_i} L_i(w_i) = \frac{1}{|D_i|} \sum_{(x,y) \in D_i} l(w_i; x, y), 
\end{equation}
\noindent while the global model is updated through weighted aggregation, \textit{i.e.},
\begin{equation}
    w_{global} = \sum_{i=1}^{N} \frac{|D_i|}{|D|} w_i,
\end{equation}
\noindent where $w_i$ represents local model parameters, $D_i$ is the local dataset at operator $i$, and $N$ is the number of participating nodes. \bluechange{Furthermore, AI-assisted successive interference cancellation can improve the coexistence of RF sensing and 5G signals in shared spectrum. Consequently, such interference-aware processing is relevant to heterogeneous sensing and communication links in LAWNs~\cite{Mohammadi2025}.}

\par Similarly, power control mechanisms have progressed from basic power budget management to intelligent energy optimization frameworks that are crucial for extending the operational capabilities of energy-constrained aerial platforms. Initially, traditional power control relied on simple signal-to-noise ratio thresholds, but RL agents have transformed power management by learning optimal allocation policies that dynamically balance communication quality with energy efficiency requirements~\cite{Chu2023}. Subsequently, multi-objective optimization approaches demonstrated superior performance in extending mission duration and operational range by jointly considering battery constraints, channel conditions, and interference mitigation requirements~\cite{Sun2021}.

\par Simultaneously, trajectory optimization has advanced from simple waypoint planning to sophisticated AI-driven frameworks that jointly optimize flight paths with communication requirements and sensing objectives critical to LAWN mission success. Initially, classical optimization techniques such as genetic algorithms and particle swarm optimization provided baseline capabilities for exploring large solution spaces while considering basic constraints, including airspace restrictions and fuel limitations. In such methods, potential solutions are represented as individual $\mathbf{x} = [x_1, x_2, \ldots, x_n]$. These individuals encode flight paths or resource allocation strategies. The algorithm evaluates each solution using a fitness function as follows:
\begin{equation}
    f(\mathbf{x}) = f_{obj}(\mathbf{x}) - \lambda \sum_{i=1}^m \max(0, g_i(\mathbf{x})), 
\end{equation}
\noindent where $f_{obj}(\mathbf{x})$ represents the primary mission objective. The constraints $g_i(\mathbf{x})$ include airspace restrictions and safety requirements. The penalty coefficient $\lambda$ enforces constraint compliance by reducing fitness when violations occur. Following this, such methods evolve solutions through heuristic operations. For instance, the genetic algorithm evolves solutions through three main operations, \textit{i.e.},
\begin{equation}
    \mathcal{P}^{(t+1)} = \text{Selection}(\text{Crossover}(\text{Mutation}(\mathcal{P}^{(t)}))), 
\end{equation}
\noindent where $\mathcal{P}^{(t)}$ denotes the population at generation $t$. Selection operations choose high-fitness individuals for reproduction. Crossover operations combine genetic information from two parent solutions. Mutation operations introduce random changes with probability $p_m$. This mutation process maintains population diversity and prevents premature convergence to local optima.

\par Recently, deep RL frameworks demonstrated superior performance in learning trajectory policies that dynamically adapt to changing environmental conditions, network performance requirements, and mission priorities~\cite{Wei2023}. Furthermore, multi-agent coordination algorithms have enabled swarm-level trajectory optimization where individual UAV paths are jointly planned to maximize network coverage, minimize interference, and ensure collision avoidance~\cite{Yu2025}. The optimization process is similar to Section~\ref{sssec: AI for CN}. Consequently, modern AI-driven trajectory optimization now supports real-time adaptive planning that responds to dynamic airspace conditions, emergency scenarios, and evolving mission requirements while maintaining compliance with regulatory constraints.

\subsubsection{AI for Network Security and Privacy}

\par \bluechange{Related studies have explored explainable AI and soft-computing methods for privacy protection in cyber-physical systems~\cite{Gan2025}. For LAWNs, security design must additionally address the physical accessibility of mobile platforms, omnidirectional wireless signal interception risks, and the distributed nature of aerial operations~\cite{Li2024b}.} Moreover, AI-driven security systems excel at operating under the strict computational and energy constraints of aerial platforms while delivering real-time threat detection and autonomous response capabilities essential for maintaining operational integrity in dynamic 3D environments. 

\par Similarly, intrusion detection systems have progressed from signature-based pattern matching to intelligent deep learning frameworks that excel at processing the multi-modal data streams characteristic of LAWN operations. Initially, traditional intrusion detection relied on predefined attack signatures, but supervised learning approaches have transformed threat classification by identifying sophisticated attack patterns including denial-of-service attacks, RF jamming attempts, and data injection attacks across heterogeneous LAWN platforms. Subsequently, deep learning models demonstrated superior performance in processing complex multi-modal data including network traffic characteristics, RF signal features, and behavioral patterns to detect advanced persistent threats that evade conventional detection methods~\cite{Quadar2024}. Moreover, federated learning frameworks have proven particularly valuable for LAWN security by enabling collaborative threat intelligence sharing among multiple operators without compromising sensitive operational information. Additionally, recent developments in graph neural networks and attention mechanisms have enhanced intrusion detection by modeling complex attack propagation patterns across distributed LAWN architectures~\cite{Lin2025}. \revise{Beyond AI-enabled intrusion detection, physical-layer security provides
a complementary defense for A2A and A2G ISAC links. Their strong LoS
propagation improves connectivity and sensing coverage, but also
increases exposure to eavesdropping, jamming, and sensing-information
leakage. A2G links are mainly threatened by ground-based adversaries,
whereas A2A links must additionally account for mobile aerial
eavesdroppers and potentially untrusted swarm members. Representative
countermeasures include secure beamforming, artificial noise,
cooperative jamming, trajectory adaptation, and channel-based secret
key generation. In ISAC systems, these mechanisms should jointly
balance secrecy performance with sensing accuracy and communication
reliability.}

\par \bluechange{Simultaneously, privacy-aware optimization in cloud-based IoT systems shows that resource management can jointly account for operational efficiency and data protection. For LAWNs, this balance is particularly important because distributed aerial platforms exchange operational data while operating under stringent energy and computation constraints. Accordingly, privacy-aware coordination should limit unnecessary information exposure while preserving the data exchange required for collaborative control and resource allocation~\cite{Chen2022}.}

\subsection{Emerging AI Paradigms for LAWN}

\par Rapid advances in AI are introducing new paradigms that promise to further improve the LAWN functionalities, thereby enabling more sophisticated reasoning, content generation, and human-AI interaction. These emerging AI technologies address limitations of current approaches while opening new possibilities for autonomous aerial systems that can understand, reason about, and communicate in natural language. In the following, we introduce and discuss some novel AI paradigms, which could likely define the next generation of LAWNs.

\subsubsection{Generative Models}

{
\par \bluechange{Building on the generative AI fundamentals introduced in Section~\ref{sec:ai-fundamentals}, generative models have been used to synthesize residential energy data in a related application domain~\cite{Chen2025}. In addition, generative data augmentation has improved long-range visual detection when labeled observations are limited~\cite{Chang2025}. These examples demonstrate the ability of generative models to learn complex data distributions from limited observations, which is particularly relevant to data-scarce LAWN operations.}}

{
\par As a prominent example of emerging generative models, diffusion-based architectures (introduced in Section~\ref{sec:ai-fundamentals}) are increasingly deployed for generating high-fidelity synthetic sensor data that are essential for robust LAWN perception systems.} \bluechange{In particular, diffusion models have improved the resolution of sparse satellite LiDAR measurements~\cite{Ramirez–Jaime2024}. Subsequently, score-based generative models have provided stable mechanisms for reconstructing complex data distributions from incomplete observations. Furthermore, conditional diffusion models have been investigated for wireless data reconstruction~\cite{Letafati2025}. Consequently, diffusion policies have also supported generalizable autonomous navigation under varying environments~\cite{Hu2025}.}

\par Simultaneously, flow-based models are emerged as powerful tools for uncertainty quantification and risk assessment that are crucial for safe LAWN operations in unpredictable low-altitude environments. Initially, normalizing flows provided basic capabilities for modeling complex probability distributions of environmental conditions and system parameters with tractable likelihood computation~\cite{Zhong2022}. Subsequently, continuous normalizing flows demonstrated superior performance in handling temporal dynamics and enabling accurate prediction of system evolution over time with quantified confidence bounds. Moreover, recent developments in neural ordinary differential equations and augmented normalizing flows have enhanced the ability to model complex spatial-temporal dependencies in multi-UAV network scenarios~\cite{Peng2025}. Additionally, invertible neural networks now enable precise uncertainty propagation through complex LAWN decision-making pipelines, supporting risk-aware trajectory planning and resource allocation. 



\par Complementarily, building on the VAE fundamentals discussed in Section~\ref{sec:ai-fundamentals}, emerging VAE variants are demonstrated particular promise for complex LAWN data processing. In particular, $\beta$-VAEs and vector quantized VAEs (VQ-VAEs) proved particularly effective for LAWN applications by learning disentangled representations that separate different factors of variation in flight data. These representations enable more interpretable and controllable generation of synthetic operational scenarios~\cite{Murad2025}. \bluechange{Furthermore, VAE-based representations have been used to reduce data dimensionality in cyber-physical forecasting tasks. Consequently, latent-space modeling may offer a compact way to represent high-dimensional LAWN operational data~\cite{Kaur2021}.}

\subsubsection{Large Language Models}

\par In particular, LLMs and foundation models represent an emerging paradigm for natural language interfaces and high-level reasoning in LAWNs, promising to revolutionize human-LAWN interaction and autonomous mission planning~\cite{Yin2025}. In particular, these AI systems enable UAVs and LAWNs to understand complex mission objectives specified through natural language commands, reason about intricate operational scenarios involving multiple constraints and objectives, and communicate findings and decisions in human-interpretable formats that enhance operational transparency. Moreover, LLMs serve as intelligent supervisory interfaces between human operators and complex LAWN systems, dramatically reducing the technical expertise required for mission planning, system configuration, and real-time operational intervention.


\par Similarly, collaborative reasoning and multi-agent coordination capabilities leverage LLMs' natural language processing abilities to facilitate seamless coordination among heterogeneous LAWN platforms. In particular, LLM integration enables natural language-based communication and negotiation among UAVs with different capabilities and operational objectives. Subsequently, LLM-mediated resource conflict resolution has demonstrated superior performance in coordinating task allocation and enabling dynamic mission replanning when operational conditions change or unexpected events occur~\cite{Zhang2025c}. Moreover, natural language interfaces now allow human operators to intuitively modify mission parameters, add new objectives, or intervene in autonomous operations without requiring detailed knowledge of underlying system architectures or programming languages. Additionally, recent developments in multi-modal LLMs have enhanced coordination capabilities by integrating visual and sensor data with natural language reasoning for more comprehensive situational awareness~\cite{Cao2025}.

\par Likewise, LLM-based knowledge integration and reasoning enable comprehensive contextual understanding by incorporating diverse information sources into LAWN decision-making processes. \bluechange{In particular, task and spatial information has been organized within a unified representation for UAV swarm decision-making~\cite{Zhu2025}. This type of structured operational knowledge can be incorporated into LLM reasoning through retrieval mechanisms.} Subsequently, vector database integration and semantic search capabilities have enhanced LLMs' ability to identify relevant information from vast operational datasets and sensor streams~\cite{Yao2024a}. Furthermore, chain-of-thought and tree-of-thought reasoning methodologies have proven particularly valuable for LAWN applications by enabling LLMs to provide detailed explanations for their operational decisions, improving transparency and building operator trust in autonomous systems. Consequently, modern LLM-powered knowledge integration now supports comprehensive situational awareness and evidence-based decision-making across complex LAWN operational scenarios~\cite{Xiang2024}.

\par \bluechange{Simultaneously, LLM-driven code generation and coordination provide flexible interfaces for adapting system behavior to operational requirements. In particular, LLMs have been used to translate natural-language descriptions into executable safety-critical simulation scenarios~\cite{Zhang2024a}. Subsequently, similar interfaces may support mission configuration and scenario construction for LAWNs. Moreover, LLMs have been combined with RL to improve communication-efficient multi-agent cooperation~\cite{Su2025}. Consequently, this integration may support adaptive coordination among heterogeneous aerial platforms.}

\subsection{Lessons Learned}

\par The exploration of AI technologies for LAWNs reveals several critical insights that guide the practical deployment and future development of intelligent aerial networks. Fundamentally, the synergistic integration of discriminative and generative AI paradigms emerges as essential rather than optional, where discriminative models provide real-time decision-making and control capabilities while generative models enable predictive adaptation and synthetic scenario generation for robust training. Moreover, the progression from classical machine learning methods to advanced deep learning architectures demonstrates a clear trajectory toward increasingly sophisticated yet computationally efficient solutions that respect the stringent constraints of aerial platforms. Additionally, the successful deployment of AI in LAWNs requires careful consideration of the trade-offs between model complexity and operational feasibility, where lightweight architectures such as MobileNet for vision tasks and state-space models like SpectralMamba often outperform larger models in resource-constrained environments. Furthermore, the incorporation of attention mechanisms across both discriminative and generative frameworks consistently improves performance by enabling models to focus on the most relevant features in the complex, multi-modal data streams characteristic of low-altitude operations.

\begin{figure*}
  \centering
  \includegraphics[width=\textwidth]{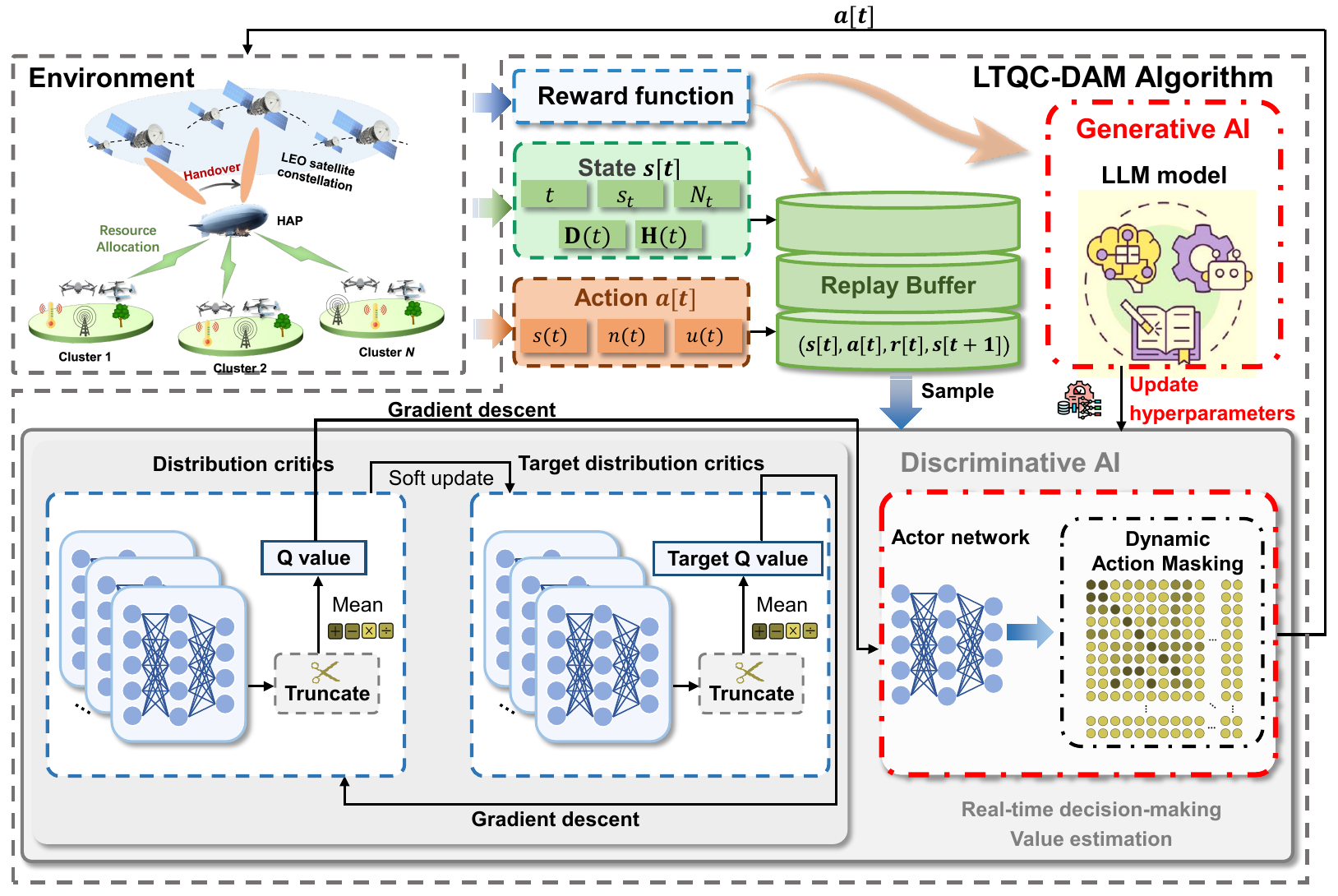}
  \caption{\bluechange{Framework of the LTQC-DAM algorithm for LAWN systems, newly illustrated based on our previous work~\cite{Li2025a}. The framework integrates a discriminative TQC agent for real-time decisions and a generative LLM meta-controller for adaptive hyperparameter tuning. Dynamic action masking prunes invalid satellite selections, while the LLM analyzes training dynamics to ensure stable learning in the non-stationary LAWN environment.}}
  \label{fig:algorithm_framework}
\end{figure*}


{
%
\section{Intelligent Signal Processing Integration: A Case Study of AI-Driven Multi-tier LAWNs}

\par \bluechange{To demonstrate the synergistic potential of discriminative AI, generative AI, and signal processing techniques in practical LAWN deployments, this section presents a case study based on our previous work~\cite{Li2025a}.} The considered multi-tier architecture integrates LEO satellites, HAPs, and low-altitude users, providing a representative scenario that captures several key challenges of LAWN operations, including mobility, dynamic topologies, heterogeneous link management, and low-altitude propagation effects. Moreover, within this architecture, discriminative and generative AI complement robust signal processing methods in channel characterization, resource allocation, link selection, and network adaptation.

\subsection{Scenario Description}

\par \bluechange{The case study considers the multi-tier satellite, HAP, and user architecture in~\cite{Li2025a}.} Specifically, this architecture provides a concrete implementation of the altitude-based layers and functional planes introduced above. The LEO satellites operate in multiple orbital planes to provide wide-area coverage, while the HAPs serve as intermediate relay nodes connecting the satellite segment with geographically distributed low-altitude users through OFDM-based communications~\cite{Elmahallawy2024}.

\par The architecture employs FSO links between LEO satellites and HAPs to utilize the available bandwidth and favorable propagation conditions at higher altitudes. RF links are used between HAPs and low-altitude users because they are generally less sensitive to weather conditions and atmospheric disturbances near the ground~\cite{Fan2025}. However, the resulting hybrid network is highly dynamic. The rapid movement of LEO satellites continuously changes link availability and network topology, leading to intermittent connectivity and frequent handovers. Since traffic is relayed across both FSO and RF segments, satellite selection and resource allocation in the FSO segment also influence the service conditions of the RF segment. Although HAPs move more slowly, changes in their positions affect link distances, channel quality, and coverage. At the low-altitude tie, buildings and terrain may cause aerial links to alternate between LoS and NLoS conditions, further complicating connectivity management and resource allocation.

\par These variations create temporal and cross-tier dependencies because each decision affects not only the current transmission state but also subsequent connectivity, resource availability, and handover options. Static analytical methods are difficult to apply directly. Model-based algorithms may require repeated optimization whenever the topology or channel conditions change, and considerable computation may be spent evaluating actions that are unavailable or unsuitable in the current state. Their performance may also decline when actual mobility patterns and propagation conditions differ from the assumptions used in the system model~\cite{Cao2025}. An integrated approach is therefore considered in which signal processing models characterize the channels and define feasible operating constraints, discriminative AI makes online resource allocation and handover decisions, and generative AI adjusts selected training hyperparameters according to reward trends and training progress.

\subsection{Synergistic AI-Signal Processing Framework}

\par \bluechange{In this case study, we organize the signal processing and learning components of LTQC-DAM~\cite{Li2025a} into an AI-enhanced signal processing framework for the considered multi-tier LAWN.} As illustrated in Fig.~\ref{fig:algorithm_framework}, signal processing models characterize the FSO and RF links, formulate the communication objectives, and define the physical and resource constraints. Moreover, within this model-based environment, a discriminative AI module learns online decision policies, while a generative AI module adjusts selected training hyperparameters according to the observed learning process.

\par First, the signal processing models describe the physical relationships among network topology, channel conditions, transmission resources, and system performance. Based on the positions of LEO satellites, HAPs, and low-altitude user clusters, the models determine the available satellite-to-HAP FSO links and HAP-to-user RF links at each time slot. They also provide the channel gains, achievable transmission rates, satellite visibility sets, and handover states required by the subsequent decision process. Let $\boldsymbol{N}$, $\boldsymbol{S}$, and $\boldsymbol{U}$ denote the subcarrier allocation, satellite selection of LAWNs, and low-altitude user association decisions, respectively. The bi-objective problem is formulated as
\begin{equation}
\underset{\boldsymbol{N},\boldsymbol{S},\boldsymbol{U}}{\mathrm{max}}
\quad
F=\left(f_1,-f_2\right),
\end{equation}
\noindent where $f_1$ denotes the downlink transmission rate and $f_2$ denotes the satellite handover frequency of LAWNs. Moreover, the signal processing models map each decision into its corresponding communication performance while defining the physical and resource constraints of the optimization problem.

\par In particular, the selected satellite needs to remain visible during each time slot, which is given by
\begin{equation}
s_t \in \mathcal{L}_t,
\quad \forall t \in \mathcal{T},
\end{equation}
\noindent where $s_t$ is the selected satellite at time $t$, $\mathcal{L}_t$ is the set of visible satellites, and $\mathcal{T}$ is the set of time slots. Likewise, the subcarrier allocation needs also to satisfy the constraint as follows. 
\begin{equation}
\sum_{i=1}^{N_C} n_i(t)=N_S,
\quad \forall t \in \mathcal{T},
\end{equation}
\noindent where $n_i(t)$ is the number of subcarriers allocated to cluster $i$, $N_C$ is the number of user clusters, and $N_S$ is the total number of available subcarriers. These models and constraints provide the state information, objective values, and feasibility conditions used to construct the learning environment.

\par \bluechange{Second, the discriminative AI module formulates the sequential optimization problem as a Markov decision process and employs TQC with dynamic action masking to learn the control policy~\cite{Li2025a}.} The state is constructed from the network information provided by the signal processing models, including channel conditions, satellite visibility, available resources, network topology, and satellite selected in the previous time slot.

\par The action at time slot $t$ is defined by the original decision variables, including subcarrier allocation, satellite selection, and user association, respectively. Since the visible satellite set changes over time, dynamic action masking restricts the satellite selection and prevents the policy from selecting unavailable satellites. Moreover, the reward combines the aforementioned optimization objectives with penalties for constraint violations. 
\begin{equation}
r_t =
\alpha_1 f_1(t)
-\alpha_2 f_2(t)
-\lambda P_{\mathrm{C}}(t),
\end{equation}
\noindent where $f_1(t)$ and $f_2(t)$ denote the normalized downlink transmission rate and satellite handover conditions at time $t$, respectively. Moreover, the coefficients $\alpha_1$ and $\alpha_2$ specify the relative importance of the two objectives. In addition, $P_{\mathrm{C}}(t)$ denotes the penalties associated with violations of constraints, while $\lambda$ is the corresponding penalty coefficient. Note that each penalty is zero when its associated constraint is satisfied and positive otherwise.

\par Based on these definitions, TQC learns a joint policy for satellite selection, user association, and subcarrier allocation that balances transmission rate, handover frequency, and constraint satisfaction.

\par \bluechange{Finally, the generative AI module introduces an LLM-based meta-controller that periodically adjusts selected hyperparameters according to recent training information~\cite{Li2025a}.} The update process is expressed as
\begin{equation}
\Theta_{e+\Delta e} = \mathcal{F}_{\text{LLM}}(\Theta_e, \mathcal{H}_e, \mathcal{P}_e),
\end{equation}
\noindent where $\Theta_e = \{\theta_1^e, \theta_2^e, \ldots, \theta_{N_\theta}^e\}$ represents the set of hyperparameters at training iteration $e$, $\mathcal{H}_e = \{r[e-k], r[e-k+1], \ldots, r[e]\}$ represents a window of $k$ recent episode rewards, and $\mathcal{P}_e = {e}/{E}$ represents the normalized training progress with $E$ being the total number of training episodes. Furthermore, the LLM meta-controller analyzes reward patterns, training dynamics, and learning progress through structured prompting, \textit{i.e.,}
\begin{equation}
\begin{aligned}
\text{prompt} = 
\begin{bmatrix}
\Theta_e &= \{\theta_1^e, \theta_2^e, \ldots, \theta_{N_\theta}^e\} \\
\mathcal{H}_e &= \{r[e-k], r[e-k+1], \ldots, r[e]\} \\
\mathcal{P}_e &= \frac{e}{E}
\end{bmatrix}.
\end{aligned}
\end{equation}
\noindent As such, the selected parameters such as learning rates, entropy coefficients, and exploration schedules can be adaptively tuned. These parameters are constrained within predefined ranges, which are given by
\begin{equation}
\theta_i^{e+\Delta e} \in [\theta_i^{\text{min}}, \theta_i^{\text{max}}].
\end{equation}

\noindent Following this, the updated hyperparameters are returned to the TQC-DAM learner and used during the next training stage. Different from the discriminative policy, which makes network decisions at each time slot, the generative AI module operates periodically and focuses on the adaptation of the learning process.

\par In summary, the three components perform different but connected functions. Signal processing models describe the FSO and RF channels, calculate communication performance, and define the feasible decision space. Moreover, the discriminative AI module uses this information to make online satellite selection and resource allocation decisions under time-varying network conditions. In addition, the generative AI module observes reward trends and training progress, and periodically adjusts the hyperparameters of the discriminative learner. Through this interaction, the signal processing models provide physical guidance, LTQC-DAM handles network-level decisions, and the LLM meta-controller supports training adaptation. As such, they provide a learning-based optimization approach to the temporal and cross-tier optimization problem considered in this tutorial case study.

\begin{figure}[t]
  \centering
  \includegraphics[width=3.5in]{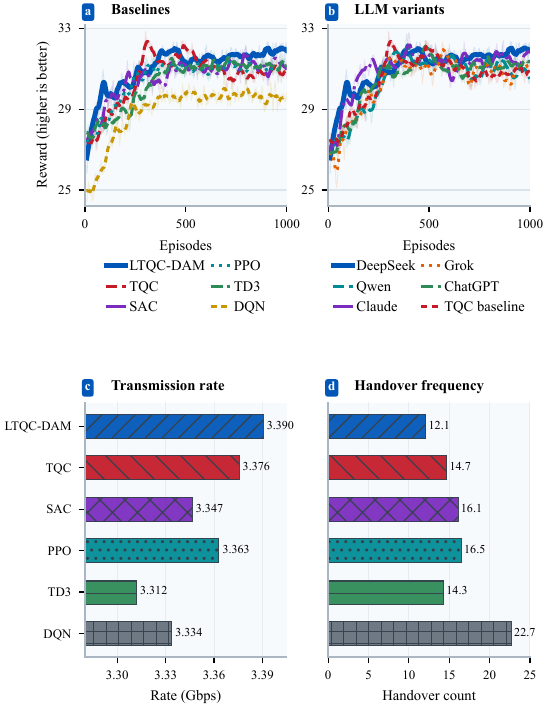}
  \caption{\bluechange{Training and optimization-objective performance of LTQC-DAM, replotted using the data reported in~\cite{Li2025a}. (a) reward comparison with algorithm baselines; (b) reward comparison across LLM-guided variants; (c) downlink transmission rate; (d) satellite handover frequency.}}
  \label{fig:performance_comparison}
\end{figure}

\begin{figure}[t]
    \centering
    \includegraphics[width=3.5in]{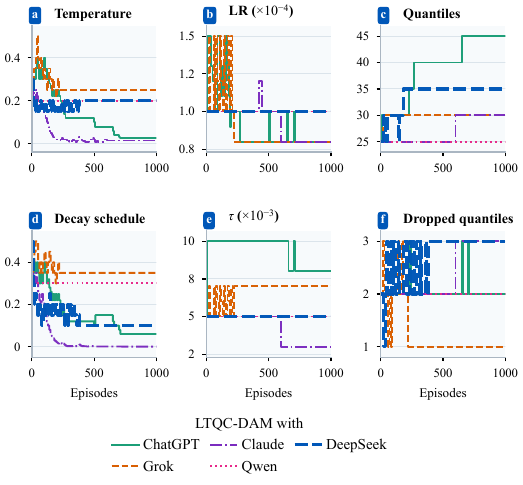}
    \caption{\bluechange{Hyperparameter adaptation patterns across different LLMs during training, replotted using the data reported in~\cite{Li2025a}.}}
    \label{fig:llm_comparison}
\end{figure}

\subsection{Evaluation Results}

\par \bluechange{Based on~\cite{Li2025a}, we compare LTQC-DAM against several baseline methods, including SAC, TD3, PPO, DQN, and standard TQC algorithms.} Moreover, we evaluate different LLM models, including DeepSeek, Qwen, Claude, ChatGPT, and Grok to assess the impact of generative AI components on system performance.

\par \bluechange{The convergence results of our previous work~\cite{Li2025a} show that LTQC-DAM exhibits faster convergence and achieves higher final rewards compared to baseline algorithms, as illustrated in Fig.~\ref{fig:performance_comparison}(a). Furthermore, as shown in Fig.~\ref{fig:performance_comparison}(b), while the standard TQC algorithm converges rapidly, it shows fluctuations in later training stages due to static hyperparameter settings. Moreover, LLM-guided hyperparameter optimization enables stable convergence by adaptively tuning critical parameters throughout the training process, addressing the challenge of parameter sensitivity in dynamic satellite and low-altitude environments.}

\par \bluechange{Moreover, the performance results of our previous work~\cite{Li2025a} show that LTQC-DAM outperforms all benchmark algorithms across both optimization objectives. As illustrated in Figs.~\ref{fig:performance_comparison}(c) and 7(d), LTQC-DAM achieves the lowest satellite handover frequency while maintaining the highest downlink transmission rate among the evaluated algorithms. Additionally, the algorithm reaches stable performance within 400-500 training episodes, representing acceleration compared to baseline methods.}
}

{
\par In addition, the hyperparameter results of our previous work~\cite{Li2025a} show that DeepSeek generally makes smaller and more gradual adjustments, as illustrated in Fig.~\ref{fig:llm_comparison}. By contrast, several other models introduce larger or oscillatory changes, particularly during the later training stages. For example, DeepSeek keeps the learning rate relatively stable and adjusts multiple hyperparameters in a coordinated manner. When increasing the entropy temperature to encourage exploration, DeepSeek also adjusts the number of truncated quantiles to maintain a more conservative critic target. By contrast, frequent or inconsistent changes observed for some other LLMs may disturb the policy update process and contribute to fluctuations during training.

\par These observations suggest that LLM-guided hyperparameter adaptation benefits from moderate updates that account for both the current training stage and the interactions among different parameters. Under the considered settings, DeepSeek is more consistent with this adjustment pattern, which may partly explain its smoother convergence and better final performance. Moreover, its ability to model dependencies across the training history may support more coordinated recommendations, although the relationship between the internal LLM architecture and optimization performance requires further investigation.
}






\section{Research Challenges and Open Issues}
In the previous sections, we have reviewed signal processing foundations, alongside various AI techniques~\cite{wu2025low}. These discussions connect to a growing body of work on LAWN architectures and intelligent ISAC/6G systems that emphasize tightly coupled communication, sensing, and control in low-altitude airspace. While promising, the field remains in early stage, many foundational questions in architecture design, signal processing–AI co-design, safety and security, and experimental validation are still open. Motivated by both the opportunities and the gaps highlighted so far, this section outlines key research challenges and open issues that must be addressed for LAWNs to evolve into dependable, AI-native infrastructure for the intelligent skies.

\subsection{Architecture-Level Challenges for Large-Scale LAWNs}
\subsubsection{Cross-plane interfaces and abstractions}
Complementing altitude layers, the LAWN architecture introduces four functional planes, exchanging state and reconfiguring resources to co-optimize safety, efficiency, and mission performance. While this decomposition clarifies roles, it raises unresolved questions about what information should flow across planes, at which timescales, and via which standardized interfaces. For example, the data plane may need real-time risk or blockage indicators from the sensing plane and airspace/corridor updates from the control plane, while the computing/intelligence plane must decide where to place optimization and learning functions under tight latency and reliability constraints. 
Designing cross-plane abstractions and hierarchical control loops that enable joint optimization without instability or excessive overhead remains a key architectural challenge.

\revise{\subsubsection{Scalability and multi-operator coexistence}
Scaling LAWNs from small-scale trials to city- or nation-wide
deployments introduces fundamental signaling, interference, and
computational bottlenecks. Dense aerial nodes generate frequent
handover, topology-update, sensing-exchange, and coordination messages,
while LoS-dominated propagation enlarges the interference footprint
across cells and altitude layers. Centralized optimization may therefore
become computationally prohibitive and vulnerable to outdated global
state information. Moreover, coexistence among regulators, UTM/U-space
providers, mobile operators, and vertical users requires interoperable
control and security policies \cite{hamissi2023survey}. Hierarchical management,
localized information exchange, distributed processing, and
event-triggered coordination are promising, but their scalability,
stability, and performance losses relative to centralized designs remain
open questions.}

\subsection{Signal Processing in Multi-Functional LAWNs}
\subsubsection{Non-stationary 3D channels and interference-limited operation}
Low-altitude channels are highly non-stationary due to 3D mobility, exhibiting intermittent LoS/NLoS transitions and and blockage dynamics \cite{wang2025stackelberg}. Signal processing must therefore enable environment- and trajectory-aware channel modeling, fast prediction, and robust estimation at swarm scale. Meanwhile, LoS-dominated links create strong inter-cell and A2A interference, calling for advanced PHY mechanisms such as 3D beamforming, interference-aware pilot design, and coordinated transmission in vertically extended, rapidly changing topologies.

\subsubsection{Multi-functional waveform and receiver co-design across planes}
LAWNs require the same radios to simultaneously support payload delivery, safety-critical control, and sensing, often under shared spectrum and hardware constraints \cite{wang2025generative}. This motivates joint waveform-pilot-receiver co-design that explicitly trades communication performance, sensing fidelity, and control-loop stability under tight power, size, and computation budgets. A further challenge is to retain compatibility with heterogeneous standards and legacy infrastructure, enabling multi-functionality without spectrum fragmentation or duplicated hardware.

\subsection{AI and Machine Learning Challenges in LAWNs}
\subsubsection{Safe and sample-efficient multi-agent learning under coupled objectives}
RL/MARL are promising for trajectory planning, coverage, and resource allocation, but operational LAWN deployment faces three coupled barriers \cite{wang2025security}. First, data efficiency is critical while safety constraints severely limit trial-and-error exploration. Second, objectives are inherently multi-dimensional, yet many designs rely on ad-hoc reward shaping that weakly enforces constraints. Third, the environment is non-stationary and partially observable due to evolving channels, traffic, and airspace rules. Developing constrained, risk-aware, and model-based/offline RL methods that learn from limited data while scaling to large swarms and multi-tier architectures remains an open problem.

\subsubsection{Resource-aware learning with foundation models in the loop}
AI in LAWNs must run across heterogeneous, resource-limited platforms spanning onboard compute, edge servers, and cloud backends, under strict latency and energy budgets. This motivates distributed learning that accounts for link variability, non-IID data, and dynamic agent participation. In parallel, foundation models are increasingly envisioned as supervisory planners for intent translation, mission planning, and algorithm selection, raising additional challenges, such as partitioning and compressing models for edge deployment, verifying suggestions against safety and regulatory constraints, and coordinating updates without destabilizing lower-level control. Designing hierarchical, resource-aware learning architectures that couple lightweight onboard policies, collaborative edge learning, and verifiable foundation-model guidance is therefore central to LAWN autonomy.

\subsection{Security in Open Low-Altitude Airspace}
\subsubsection{Cross-layer safety, security, and trustworthy operation}
\revise{The tight coupling of communication, sensing, control, and AI enlarges
the attack surface of LAWNs beyond conventional wireless security
\cite{wang2022security}. Jamming and eavesdropping may disrupt A2A/A2G links,
spoofed sensing or localization measurements may corrupt situational
awareness, and falsified control commands may directly affect flight
safety. Learning-enabled functions are additionally vulnerable to data
poisoning, adversarial observations, model extraction, and unreliable
foundation-model outputs. These threats may propagate across functional
planes and amplify one another. Practical protection therefore requires
authenticated sensing and control information, physical-layer security,
runtime anomaly detection, trustworthy AI, and fail-safe control
mechanisms. A key unresolved issue is how to provide such guarantees
under stringent latency, energy, and onboard-computing constraints.}

\subsubsection{Resilience under failures, uncertainty, and extreme events}
Beyond prevention, LAWNs must remain operational under outages, congestion, cyber-physical attacks, and extreme weather \cite{wang2025toward}. This motivates redundancy across altitude layers and planes and adaptive control/AI policies that reconfigure connectivity while keeping risk bounded. Signal processing and learning should explicitly account for uncertainty in channels, sensing, and traffic, enabling graceful degradation rather than brittle behavior. Providing quantitative robustness guarantees, e.g., minimum service continuity is an important direction toward deployable LAWNs.

\subsection{Experimentation, Benchmarking, and Standardization for LAWNs}

\subsubsection{Open datasets, simulators, and field testbeds}

\revise{Several representative resources are available for experimental LAWN research. AERPAW supports programmable wireless experiments using autonomous UAVs and provides air-to-ground channel, spectrum, localization, and telemetry datasets\cite{AERPAW}. AirSim supports software- and hardware-in-the-loop evaluation for aerial perception, navigation, and control\cite{AirSim}, whereas 5G-LENA enables full-stack simulation of 3GPP NR networks\cite{FiveGLENA}. DeepMIMO further provides reproducible ray-tracing-based MIMO channel datasets for data-driven wireless research\cite{DeepMIMO}. Nevertheless, a major obstacle to real-world LAWN deployment is the gap between simulation assumptions and operational environments. Existing studies often adopt simplified channel, mobility, traffic, sensing, and energy models that do not fully capture hardware impairments, weather effects, urban blockage, asynchronous operation, or regulatory restrictions. Meanwhile, representative multi-modal field datasets and large-scale testbeds remain limited, making reproducibility and fair comparison difficult. Future evaluation should therefore combine high-fidelity digital twins, hardware-in-the-loop platforms, controlled flight tests, and progressively scaled field trials. Standardized scenarios, metrics, interfaces, and baselines are also needed to assess not only algorithmic
performance, but also robustness, interoperability, implementation cost, and safety under realistic operating conditions.} 

\subsubsection{Standardization, interoperability, and regulation--technology co-evolution}
Deployable LAWNs demand interoperability across vendors and operators, requiring well-defined interfaces and standards that reflect AI-native, multi-plane operation. While cellular, NTN, and UTM/U-space standards offer building blocks, they only partially capture the tight coupling of multiple functionalities in LAWNs. Open issues include extending these frameworks with LAWN-specific functions, data models, and performance indicators while remaining implementable across platforms from small UAVs to HAPS and satellites. In parallel, regulation and technology must co-evolve via digital twins, testbeds, and staged deployments so that safety, security, and privacy requirements translate into concrete protocol and system guidelines and can be refined as operational evidence accumulates.

\section{Conclusions}
This tutorial framed the LAWN as a unifying paradigm for communications, sensing, and control in safety-managed 3D airspace. We reviewed its interdisciplinary roots and proposed an architecture based on altitude-stratified layers and functional planes to structure payload transport, command-and-control/Remote ID, perception, and distributed computing, while aligning with evolving UTM/U-space concepts and cellular/non-terrestrial networking. Building on this foundation, we surveyed core signal processing problems, including joint performance metrics spanning reliability, sensing accuracy, and control stability, and the design of ISAC-capable waveforms, receivers, and multi-modal localization/tracking. We highlighted multi-function co-design, where beamforming, resource allocation, and waveform design are optimized jointly for data delivery, situational awareness, and closed-loop control. We further discussed AI as a key enabler of autonomy, from perception and prediction to learning-based trajectory and resource management, and emphasized hierarchical, resource-aware deployment across onboard, edge, and cloud platforms. \bluechange{A multi-tier case study based on~\cite{Li2025a} illustrated how learning and model-assisted optimization can address handovers and dynamic link selection under mobility constraints.} Finally, we outlined open challenges in LAWNs. Overall, LAWNs sit at the nexus of 6G/NTN, ISAC, and agentic AI, and coordinated advances can enable scalable, dependable low-altitude connectivity.

\bibliography{ref}
\bibliographystyle{IEEEtran}

\end{document}